\documentclass[12pt]{iopart}
\pdfoutput=1
\usepackage{color}
\usepackage{graphicx}
\usepackage{amsmath, amsthm, amssymb}
\usepackage[normalem]{ulem}
\definecolor{Gray}{gray}{0.7}
\usepackage{citesort}
\usepackage{tabularx}
\usepackage{amsfonts}
\usepackage{mathtools}
\usepackage{tikz}
\usetikzlibrary{positioning,decorations.pathmorphing,decorations.markings,arrows}
\usepackage{physics}
\usepackage{latexsym,amsfonts,color,amsthm}
\usepackage{enumerate}
\usepackage{fancyhdr} 
\usepackage[square, numbers, comma, sort&compress]{natbib}
\usepackage{glossaries}
\usepackage{colortbl}
\usepackage[export]{adjustbox}
\usepackage[titletoc]{appendix}

\newcommand{\be}{\begin{equation}}
\newcommand{\ee}{\end{equation}}
\newcommand{\bea}{\begin{eqnarray}}
\newcommand{\eea}{\end{eqnarray}}

\begin{document}

\title[Asymptotic velocity distribution of a one dimensional binary gas
]{Asymptotic velocity distribution of a driven one dimensional binary granular Maxwell gas} 
\author{Apurba Biswas$^{1,2}$, V. V. Prasad$^{3}$, and
R. Rajesh$^{1,2}$}

\address{$^1$ The Institute of Mathematical Sciences, C.I.T. Campus, Taramani, Chennai 600113, India}
 \address{$^2$ Homi Bhabha National Institute, Training School Complex, Anushakti Nagar, Mumbai 400094, India}
\address{$^3$ Department of Physics of Complex Systems, Weizmann Institute of Science, Rehovot 7610001, Israel}
 \ead{biswas.apurba6@gmail.com, prasad.vv@weizmann.ac.il, rrajesh@imsc.res.in}

\date{\today}

\begin{abstract}
We consider the  steady states of a driven inelastic Maxwell gas consisting of two 
types of particles with scalar velocities. Motivated by experiments on bilayers where only one layer is driven, we focus on the case when only one of the two types of particles are 
driven externally, with the other species receiving energy only
through inter-particle collision. The velocity $v$ of a particle that
is driven is modified to $-r_w v+\eta$, where $r_w$ parameterises the dissipation upon the driving and the noise $\eta$ is taken from a fixed distribution. We characterize the statistics for small velocities by computing exactly the mean energies of the two species, based on the simplifying feature  that the correlation functions are seen to form a closed set of equations.
The asymptotic behaviour of the  velocity distribution for large speeds is determined for both components through a combination of exact analysis for a range of parameters or obtained numerically to a high degree of accuracy from an analysis of  the large moments of velocity.
We show that the tails of the velocity distribution for both types of particles have similar behaviour, even though they are driven differently.  For dissipative driving ($r_w<1$), the tails of the steady state velocity distribution show non-universal features and depend strongly on the noise distribution. On the other hand, the tails of the velocity distribution are exponential for diffusive driving ($r_w=1$) when the noise distribution decays faster than exponential.
\end{abstract}

\maketitle

\section{\label{sec1-Introduction}Introduction}

One of the main features that makes granular matter different from equilibrium many particle 
systems is their  inelastic collisional interaction. Along with many intriguing phenomena exhibited by 
granular matter such as jamming, phase segregation, clustering, and others~\cite{Jaeger:96,Aranson:06,Goldhirsch:93,Li:03,Corwin:05}, the dissipative interactions  give rise to non-trivial characteristics 
even in its simplest variant, namely granular gas. Granular gases are dilute collections of particles 
that move ballistically and  interact via inelastic binary collisions. One of the central questions
is the nature of the velocity distribution, in particular is it universal and if yes, what is it? 

When isolated, the granular gas cools through dissipative collisions. The total energy decays in 
time $t$ as a power law $t^{-\theta}$, where $\theta=2$ in the initial homogeneous regime~\cite{Haff:83}, and $\theta=2d/(d+2)$ in the later inhomogeneous regime when particles cluster together~\cite{Goldhirsch:93,Brey:96,Esipov:97,Ben-naim:99,Nie:02,Supravat:12,Pathak:14a,Pathak:14,shinde2007violation}. For generic initial velocities, the time dependent velocity distribution $P(v,t)$, at large times, has a universal tail, $\ln P(v,t) \propto - v^{2/\theta} t$, for large speeds $v$~\cite{Ben-naim:99,Nie:02,Pathak:14}, in both the homogeneous as well as the inhomogeneous regimes.

When driven by external input of energy, the granular gas approaches a time independent  steady state at large times. The steady state velocity distribution $P(v)$ generically has a stretched exponential form $P(v) \sim \exp(-a v^\beta)$ for large speeds $v$, where $a$ is a constant. Though determining the value of the exponent $\beta$ has been the subject of many studies, the results are still not conclusive. Among the different systems that have been studied, the monodispersed  gas, consisting of only one type of particle, is the best studied. For this case, a number of experiments~\cite{Losert:99,Rouyer:00,Aranson:02,reis2007forcing,wang2009particle,tatsumi2009experimental,Scholz:17,vilquin2018shock} and simulations~\cite{Moon:01,gayen2008orientational,gayen2011effect,Cafiero:2002}  conclude that  the exponent $\beta$ is approximately $1.5$, and universal in the sense that it does not differ  for a wide range of densities and driving parameters. 
Theoretically, Boltzmann equation, that assumes molecular chaos such that spatial correlations may be ignored, and where driving is phenomenologically modelled by a diffusive term -- as when the system is in contact with a heat bath -- predicts $\beta=3/2$~\cite{Noije:98}. At the same time there are several other experiments which predicted a wide range of values for  $\beta$ instead of a universal form~\cite{Olafsen:99,Kudrolli:00,Blair:01,Falcon:2013,grasselli2015translational,windows2013boltzmann,wildman2009granular,Schmick:08,hou2008velocity,baxter2007temperature}. Numerical studies~\cite{puglisi1998clustering,puglisi1999kinetic,Moon:01,Vanzon:04,Vanzon:05,rui2011velocity,Prosendas:18,kang2010granular} of models of driven systems have also obtained estimates of $\beta$ that are different from $1.5$. 
Intermediate power law behaviour were also found in the case of extremal driving~\cite{kang2010granular}, wherein large momentum is imparted  with small rate to single particles. In recent work~\cite{Prasad:18,Prasad:19}, we have analysed simple models within the molecular chaos assumption, 
where noise is modelled microscopically as a discrete process, as is expected in an experiment. In these models the rate of collision is proportional to a power of the relative velocity. We showed that there are two universal regimes. One in which the velocity distribution decays as an exponential with logarithmic corrections, and other in which it decays as a gaussian with logarithmic corrections. In addition to these universal distributions, there are choices for the noise distribution for which the velocity distribution is non-universal in the sense that the tail of the distribution depends strongly on the tails of the noise distribution~\cite{Prasad:18,Prasad:19}.

The nature of steady state velocity distributions in binary gases, consisting of two different types of constituent particles, is less explored. 
Steady states of binary gases have been studied using one of two kinds of driving: one, in which both types of particles are driven~\cite{Feitosa2002,barrat2002lack,Marconi:02,wang:2003,Wang2008,JJBrey-non-eq:2009,Uecker:2009}  and the other where only particles of one type are driven~\cite{Baxter:03,baxter2007temperature,Baxter-PRL:2007,Comb-PRE:2008,Burdeau-PRE:2009,windows2013boltzmann}. 
When both types of particles are driven, the steady state velocity statistics has been analysed mostly within numerical and analytical studies. A notable feature seen in these studies have been the overpopulation near the tails of the distribution when compared to a gaussian~\cite{Pagnani-PRE:2002,Barrat-PRE:2002,Uecker:2009,Marconi:02}. In particular, Monte Carlo study of a two dimensional system of hard sphere binary gas 
has observed non-Maxwellian statistics near the tails for both bulk forcing as well as boundary driving with gravity, with lighter particles having broader tails~\cite{Pagnani-PRE:2002}. 
However,  molecular dynamics study of a two dimensional granular gas driven vertically, showed very similar velocity distributions in the horizontal direction for both the species~\cite{Barrat-PRE:2002}.
The velocity distribution depends on the nature of driving also.  Event driven simulation of a poly-dispersed system in three dimensions with momentum conserving, species independent driving~\cite{Uecker:2009} show  overpopulation of the tails for constant force  driving, in contrast with under population for   constant energy or constant velocity driving.

The scenario where only a type of particle is driven has been studied in experiments and simulations. Experiments typically
employ quasi-two-dimensional system of bilayers where the bottom layer is driven and the top layer gains energy through collisions with the bottom layer~\cite{Baxter:03,baxter2007temperature,Baxter-PRL:2007,windows2013boltzmann,Comb-PRE:2008}. 
The experiments typically find the velocity distribution  to deviate  from Maxwellian for the lower set of particles, but observe a gaussian form for the  upper layer for a large range of parameters, along with decreased correlations among top-layer particles~\cite{Baxter:03,baxter2007temperature,Baxter-PRL:2007,windows2013boltzmann}. This observation was further supported by the demonstration that the  velocity distribution of a single monomer above a vibrating bed of dimers is close to gaussian~\cite{Comb-PRE:2008}.  
Numerical simulations of the  above experimental set up  observe a deviation from gaussian as the density or mass of the top layer 
particles are increased~\cite{Burdeau-PRE:2009}, indicating  the possibility that the gaussian form for the top layer particles arises from more number of randomizing inter-layer collisions than intra-layer collisions. 

Analytical studies of driven binary systems, when only one component is driven, is lacking. The tails of velocity distributions are best studied analytically, as experimental or numerical data typically suffer from poorly sampled tails. One of the simplest models of driven granular systems that is amenable to analysis is the Maxwell model~\cite{Bobylev:00,Ben-naim:00,Baldassarri:02,Ernst:02_a,Ernst:02,Krapivsky:02,Ben-naim:02,Antal:02,Santos:03,Prasad:13,Prasad:14,Prasad:17,Marconi:02}  where the rates of collision between particles are assumed to be independent of their relative velocity. While this simplification is not consistent with ballistic gases, where the collision rates are proportional to the relative velocity, some of the qualitative results obtained in the Maxwell limit can be carried over to the more general case. 
For binary gases, where both components are driven, analysis of simple Maxwell type models in the presence of random forcing along with viscous drag shows that  the distribution near the tails is Gaussian~\cite{Marconi:02}.

In this paper, we formulate a Maxwell type model for binary gases in one dimension where only one component is driven and analyse the tails of the velocity distribution for both components for different kinds of driving schemes. There are in general two types of driving: dissipative in which during each driving event, some amount of the velocity is dissipated with additive noise, or diffusive driving where the small noise limit corresponds to a diffusive term in the Boltzmann equation. In addition, distribution of noise is also variable, as the noise does not arise as a sum of many stochastic events, but rather usually is the result of a single discrete collision with a wall. We analyse the tails for a generic driving scheme and identify universal regimes, where the tail is independent of the noise, and non-universal regimes where the tails are mostly determined by the noise. In particular, we show that the exponent $\beta$ characterising the stretched exponential decay of the velocity distributions are the same for both components, though only one component is driven.

The remainder of the paper is organised as follows. Section~\ref{sec2-The model} contains a precise definition of the model and detailed justifications for the particular form of driving that is used. In Sec.~\ref{sec3-Calculation of two point correlations}, we show that the equations for the time evolution of two-point velocity-velocity correlation functions do not involve higher order correlations. This allows an exact calculation of the mean steady state energies of the two components. The temporal dependence of the correlations are studied through Monte Carlo simulations.  Section~\ref{sec6-characteristic function} contains an exact analysis, based on characteristic functions, of the tails of the velocity distribution for diffusive driving.  In Sec.~\ref{sec5-Moment Analysis}, we determine the tails of the velocity distribution for general driving based on an exact numerical analysis of the asymptotic behaviour of large moments of the velocity. The method is compared with the exact results from diffusive driving for benchmarking. In Sec.~\ref{sec7-Analysis of tail by truncating the driving term}, we determine the tails of the velocity distribution by analysing the Boltzmann equation with driving modelled by a diffusive term. These results are compared with the exact results to determine its validity. Section~\ref{sec8-Discussions} contains a summary of results and a detailed discussion of their implications.

\section{\label{sec2-The model}The Model}

Consider a granular gas composed of two types of constituent particles $A$ and $B$ of mass $m_A$ and $m_B$ respectively. 
The number of particles of type $A$ and $B$  are $N_A$ and $N_B$ respectively with  $N_A+N_B=N$.   Particle $i$, where $i=1,\ldots, N$ and type $k$, $k \in\{A,B\}$, is characterised by a scalar velocity 
$v_{i,k}$. These velocities  evolve in time through binary collisions and external driving.  A pair of particles of type $k$ and $l$, 
where $k,l\in\{A,B\}$, collide with rate ${\lambda_{kl}}/N$. The factor $1/N$ in the collision rates ensures that the total rate of collisions between $N_k[N_k-1]/2$ 
pairs of similar type of particles and that between $N_AN_B$ pairs of different type of particles  are proportional 
to the system size $N$. During a collision, momentum is conserved, but energy is dissipated. Let $v_{i,k}$ and $v_{j,l}$ denote the pre-collision velocities and $v'_{i,k}$, $v'_{j,l}$ 
denote the post-collision velocities. Then
\begin{align}
    v'_{i,k} &= v_{i,k} - (1+r_{kl})\frac{m_l}{m_k + m_l}(v_{i,k} - v_{j,l}), \nonumber \\  
    v'_{j,l} &= v_{j,l} + (1+r_{kl})\frac{m_k}{m_k + m_l}(v_{i,k} - v_{j,l}), \qquad k,l=A, B,
    \label{collision}
\end{align}
where $r_{kl} \in [0,1]$ is the coefficient of restitution for the
collision, and $m_k$ and $m_l$ are the masses. There are three
coefficients of restitution: $r_{AA}$, $r_{BB}$, and $r_{AB}$ depending on whether the pair of colliding particles are of type \textit{AA}, \textit{BB}, or \textit{AB}. It is convenient to define
\be
\alpha_{kl}= \frac{1+r_{kl}}{2}, ~~k, l = A, B,
\label{eq:alpha}
\ee
where $1/2\leq \alpha_{kl}\leq 1$.

In addition to collisions, the system evolves through external driving. Each $A$ particle is driven at a rate $\lambda_d$. During such an event, the velocity of the driven particle is modified according to
\begin{equation}
v'_{i,A}=-r_w v_{i,A} + \eta, \quad -1<r_w \leq 1,~~~ i=1,2,\ldots, N_A,  \label{driving}
\end{equation}
where $r_w \in (-1,1]$ is a parameter and $\eta$ is noise drawn from a fixed distribution $\phi(\eta)$.   There is no compelling reason for 
$\phi(\eta)$ to be Gaussian. We assume a generic normalized stretched exponential distribution for the noise $\eta$:
  \begin{align}
        \phi(\eta) = \frac{\gamma c^{1/\gamma}}{2\Gamma(\gamma^{-1})} \exp(-c|\eta|^{\gamma}) , ~~c, \gamma >  0,   \label{noise}
    \end{align}
where $\Gamma$ is the gamma function. The distribution $\phi(\eta)$ is characterised by the exponent $\gamma$ with $\gamma=2$ corresponding to a gaussian distribution and $\gamma=1$ corresponding to an exponential distribution. Note that the limit $r_w=-1$ corresponds to random acceleration, when the momentum of the centre of mass performs a random walk, and hence the system does not reach a steady state.

Particles of type $B$ are not driven. Rather, they gain energy through collisions with \textit{A} particles. This mimics the experiments on bilayer systems~\cite{Baxter:03}, where the particles in the bottom layer are  driven through collisions with a vibrating wall (similar to $A$ particles), while the  particles in the top layer gain energy by collision with the particles of bottom layer (like $B$ particles). 

In the model, the spatial degrees of freedom have been neglected. This corresponds to the well-mixed limit where the spatial correlations between particles are ignored. In addition, we have assumed that the collision rates are independent of the relative velocity of the colliding particles. This corresponds to the so called Maxwell limit. A more realistic collision kernel would be one where collision rates are proportional to the magnitude of the relative velocity, corresponding to ballistic motion. However, the Maxwell gas is more amenable to exact analysis than models with more complicated collision kernels. In this paper,  we therefore restrict ourselves to this case.

The form of driving that has been used [see Eq.~(\ref{driving})] has several motivations. First is that the system is driven into a steady state (see Sec.~\ref{sec3-Calculation of two point correlations} for more details) for $r_w \neq -1$ unlike the case of random acceleration ($r_w=-1$) where steady state does not exist. 

Second, the form of driving may be motivated from modelling collisions of $A$ type particles with a wall. If  the wall is massive and the particle-wall collision times are assumed to be random, then  Eq.~(\ref{driving}) can be derived, where the parameter  $r_w$ is identified with coefficient of restitution  of the particle-wall  collisions. In this interpretation, $r_w \in [0,1]$~\cite{Prasad:13}.

Third, in the limit $r_w=1$, the diffusive term that is usually used to model driving in kinetic theory result can be realised~\cite{Prasad:19}. This may be argued as follows. Let $P_k(v,t)$, where $k=A,B$,  denote the probability that  a randomly chosen particle of type $k$ has velocity $v$ at  time $t$. Its time evolution is given by:
\begin{align}
           & \frac{d}{dt}P_k(v,t) = \frac{\lambda_{kk}(N_k-1)}{N} \int \int dv_1 dv_2 P_{k}(v_1, t) P_{k}(v_2, t) \delta[ ( 1-\alpha_{kk})v_1 +  \alpha_{kk} v_2-v]\nonumber \\
           & + \frac{\lambda_{k\bar{k}}N_{\bar{k}}}{N} \int \int dv_1 dv_2 P_{k}(v_1, t) P_{\bar{k}}(v_2, t) \delta[ ( 1-X_{\bar{k}})v_1 +  X_{\bar{k}}v_2-v] -\frac{\lambda_{kk}(N_k-1)}{N}  P_{k}(v, t)  \nonumber \\
           & -\frac{\lambda_{k\bar{k}} N_{\bar{k}}}{N}  P_{k}(v, t) +\delta_{k,A}\lambda_d\left[ - P_k(v, t) + \int \int d\eta dv_1 \phi(\eta) P_k(v_1, t) \delta[ -r_wv_1 + \eta - v]\right], 
           \label{pa_single}
\end{align}
where 
\begin{equation}
\bar{k}=
\begin{cases}
B,\quad   \text{if}\quad k=A,   \\
A,\quad  \text{if}\quad k=B,     \label{eq:particle label}
\end{cases}
\end{equation}
and
\be
X_{k} =\alpha_{AB} \mu_k~~\text{where }~~\mu_k=\frac{2m_{k}}{m_A+m_B}, \quad k = A, B,
\label{eq: redefine x}
\ee
with $\mu_k\in(0,2)$ and $\alpha_{AB}$ is defined in Eq.~(\ref{eq:alpha}).
While writing Eq.~(\ref{pa_single}), we have used the product measure for the joint distribution $P(v_1,v_2)=P(v_1)P(v_2)$ due to lack of correlation between velocities of different particles. This follows from the fact that pairs of particle collide at random (also see Sec.~\ref{sec3-Calculation of two point correlations} where it is shown that two-point correlations vanish). The first two terms on the right hand side of Eq.~(\ref{pa_single}) describe  gain terms due to collisions with like and unlike particles respectively. The third and fourth terms describe the loss terms due to collisions with  like and unlike particles respectively. The fifth term describe the loss and gain terms due to driving of $A$ type particles. We now focus on the  fifth term that arise from driving. Denoting it by $I_D$, we obtain
\begin{equation}
        I_D=  -\lambda_d P_A(v, t) + \lambda_d \int \int d\eta dv_1 \phi(\eta) P_A(v_1, t) \delta[ -r_w v_1 + \eta - v]. \label{driving1}
\end{equation}
        Integrating over $v_1$ and setting $r_w=1$, we get 
        \begin{align}
         I_D=  -\lambda_d P_A(v, t) + \lambda_d \int d\eta \phi(\eta) P_A(v-\eta, t), ~~r_w=1, \label{driving2}
        \end{align}
where we have used the fact that $P(v,t)$ has  the symmetry $P(v,t)=P(-v,t)$. Taylor expanding the integrand about $\eta=0$ and then integrating over $\eta$,  Eq.~(\ref{driving2}) reduces to
\begin{align}
I_D= \frac{\lambda_d}{2}\langle \eta^2 \rangle_\phi \frac{\partial^2}{\partial v^2}P_A (v,t) + \textnormal{higher order terms in $\eta$}. \label{driving3}
\end{align}
where $\langle \ldots \rangle_\phi$ denotes averaging over the noise distribution. If the higher order terms are ignored, then the driving term $I_D$ reduces to the diffusive term that is often used to model input of energy in kinetic theory~\cite{Noije:98}. It is not a priori clear when this truncation is valid. Since our model includes this limit, we will be able to test the regime of validity of this truncation.

\section{Calculation of two point correlations} \label{sec3-Calculation of two point correlations}

In the section, we study two point velocity correlation functions. 
Further, considering the steady state values of the correlations, we illustrate 
the absence of correlations in the thermodynamic limit ($N \rightarrow \infty $) for the  binary gas. Consider the different two point correlation functions:
\begin{align}
&\Sigma^A_1(t)=\frac{1}{N_A}\sum^{N_A}_{i=1} \langle v^2_{i,A}(t) \rangle, \hspace{87pt}\Sigma^B_1(t)=\frac{1}{N_B}\sum^{N_B}_{i=1} \langle v^2_{i,B}(t) \rangle,\nonumber\\
&\Sigma^{AB}_2(t)=\frac{1}{N_A N_B}\sum^{N_A}_{i=1} \sum^{N_B}_{j=1} \langle v_{i,A}(t) v_{j,B}(t) \rangle,\hspace{12pt}\Sigma^{AA}_2(t)=\frac{1}{N_A (N_A-1)}\sum^{N_A}_{i=1} \sum_{\substack{j=1 \\ j \neq i}}^{N_A} \langle v_{i,A}(t) v_{j,A}(t) \rangle, \nonumber \\
&\Sigma^{BB}_2(t)=\frac{1}{N_B (N_B-1)}\sum^{N_B}_{i=1} \sum_{\substack{j=1 \\ j \neq i}}^{N_B} \langle v_{i,B}(t) v_{j,B}(t) \rangle. \label{2 point correlations}
\end{align}
The subscript `2' denotes that the correlations are between two different particles while the correlation functions with subscript `1' measure the mean energy.

From the stochastic rules of evolution \eref{collision} and \eref{driving}, the time evolution  of the correlation functions can be obtained. They do not depend on higher order correlations, rather form a closed set of equations that may be written compactly in matrix form as
\begin{align}
\frac{d\boldsymbol{\Sigma}(t)}{dt}=~\boldsymbol{R}\boldsymbol{\Sigma}(t)+C,
\label{two-point-correlation-matrix-evolution}
\end{align}
where
\begin{align}
\boldsymbol{\Sigma}(t)&=\begin{bmatrix}
\Sigma^{A}_1(t), &
\Sigma^{B}_1(t), &
\Sigma^{AB}_2(t), &
\Sigma^{AA}_2(t), &
\Sigma^{BB}_2(t)
\end{bmatrix}^T, \\
C&=\begin{bmatrix}
\lambda_d~\sigma^2, &
0, &
0, &
0, &
0
\end{bmatrix}^T,
\end{align}
where $\sigma^2\equiv \langle\eta^2\rangle_{\phi}$ and  the matrix $\boldsymbol{R}$ is given by
{\footnotesize{
\begin{equation}
\left[\begin{array}{ccccc} 
R_{2B}\!-\!R_{1A}\!-\!R_{3B}\!-\!R_d&R_{2B}&-2 R_{2B}+R_{3B}&2R_{1A}&0\\
R_{2A}&R_{2A}\!-\!R_{1B}\!-\!R_{3A}& -2 R_{2A}+R_{3A}&0&-R_{1B}\\
\frac{R_{3A}}{2 N_A}-R_4&\frac{R_{3B}}{2 N_B}-R_4&\frac{4R_4+R_{3B}-R_{3A}}{2}&-\frac{R_{3A}}{2 N_A}+\frac{R_{3A}}{2}  &-\frac{R_{3B}}{2 N_B}+\frac{R_{3B}}{2}\\
\frac{R_{1A}}{N_A-1}&0&R_{3B}& \frac{R_{1A}}{1-N_A}\!-\!R_{3B}\!-\!\frac{R_d}{1-r_w}&0\\
0&\frac{R_{1B}}{N_B-1}&R_{3B}&0&\frac{2R_{1B}}{N_B-1}\!-\!R_{3B}\end{array}\right].
\end{equation}}}
The constants  $R_{1k},R_{2k},R_{3k}, R_4, R_d$ are functions of the collision and driving  rates as well as the coefficient of restitutions and are given by:
\begin{align}
&R_{1k}=\frac{\lambda_{kk}\alpha_{kk}(1-\alpha_{kk})(N_k-1)}{N} ,  &R_{2k}=\frac{\lambda_{AB}N_kX_k^2}{N}, \nonumber \\
&R_{3k}=\frac{2 R_{2k}}{X_k},  &R_4=\frac{\lambda_{AB}X_AX_B}{N},  \\
&R_d=\lambda_d(1-r_w^2) \nonumber.
\end{align}

We solve for the steady state values of these velocity correlation functions by setting the time derivatives in Eq.~\eref{two-point-correlation-matrix-evolution} to zero. In the thermodynamic limit  $N\rightarrow \infty$, 
with $N_A\rightarrow \infty$ and $N_B\rightarrow \infty $ such that 
 \begin{align}
 N_A=\nu_A N,~N_B=\nu_B N,~\nu_A+\nu_B=1,
 \end{align}
$\nu_A$ and $\nu_B$ being the fraction of $A$ and $B$ type particles respectively, the solutions are given by
{\small\begin{align}
&\Sigma^{A}_1 = \frac{\lambda_d \sigma^2}{\begin{aligned} 
\lambda_d(1-r^2_w)+2\nu_A \lambda_{AA}\alpha_{AA}(1-\alpha_{AA})
+\lambda_{AB}\nu_B X_B(2-X_B)-X^2_AX^2_B \nu_A \nu_B \lambda^2_{AB}\mathcal{Q}
\end{aligned} }, \label{66}\\
&\Sigma^B_1 =  \Sigma^{A}_1 \lambda_{AB} \nu_A X^2_A\mathcal{Q}, \label{67}\\
&\Sigma^{AB}_2 = \Sigma^{AA}_2=\Sigma^{BB}_2 =0\label{68},
\end{align}
where
\be
\mathcal{Q}=\frac{1}{(2-X_A) X_A \nu_A \lambda_{AB}+2\alpha_{BB}\lambda_{BB} \nu_B(1-\alpha_{BB})}.
\ee
These results show that the values of the mean kinetic energy for both types of particles are finite and that all two point correlations involving two different particles are zero. The latter
justifies the molecular chaos assumption, where joint probability distributions are split  
into product of single-point distributions.

While the steady state values of the correlations can be determined exactly, the approach to steady state cannot be determined analytically. We study the time evolution using Monte Carlo simulations, as well as direct numerical integration of Eq.~\eref{two-point-correlation-matrix-evolution} using Euler method. This acts as an additional check for the analytical calculations.

We briefly describe the Monte Carlo algorithm.  
Given a configuration of velocities at a time $t$, the system is evolved as follows. At the next time step, one of these events can occur: collisions between $AA$, $BB$, or $AB$ particles, or driving of  $A$ particles. The probabilities of these four events are 
$\lambda_{AA} N_A (N_A-1)/(2 N \mathcal{R})$, 
$\lambda_{BB} N_B (N_B-1)/(2 N \mathcal{R})$, 
$\lambda_{AB} N_A N_B/(N \mathcal{R})$, 
and
$\lambda_{d} N_A/ \mathcal{R}$, respectively, where
\be
\mathcal{R}= \frac{\lambda_{AA} N_A (N_A-1)}{2N} + \frac{\lambda_{BB} N_B (N_B-1)}{2N} + \frac{\lambda_{AB} N_A N_B}{N}+ \lambda_d N_A.
\ee
If one of the first three events were chosen, then a pair of appropriate particles are chosen at random. If the fourth event is chosen, then an $A$ particle is chosen at random. After updating the velocities of the particles,  time is incremented by $1/\mathcal{R}$.

Initially, all particles are taken to be at rest. The time evolution of the two point correlations $\Sigma_2$ for $N=100$ are shown on Fig.~\ref{figure5}(a). The data for both Monte Carlo simulations and numerical integration of Eq.~\eref{two-point-correlation-matrix-evolution} coincide. The correlations  increase with time and saturate at large times to a value different from zero, in apparent contradiction to the result obtained in Eq.~(\ref{68}). To show that these correlations vanish in the thermodynamic limit, we do finite size scaling. The data for $\Sigma^{BB}_2$ for 3 different $N$ are shown in Fig.~\ref{figure5}(b). The steady state value decreases with $N$. The data for different $N$ collapse onto a single curve when scaled as
\be
\Sigma_2(N,t)= \frac{1}{N} g(t), \label{scaling}
\ee
where the scaling function $g(t)$ is a constant for large argument. Thus, for large $N$, the correlations $\Sigma_2$ decrease to zero as $1/N$, consistent with Eq.~(\ref{68}).
\begin{figure}
\centering
		\includegraphics[width=0.95 \columnwidth]{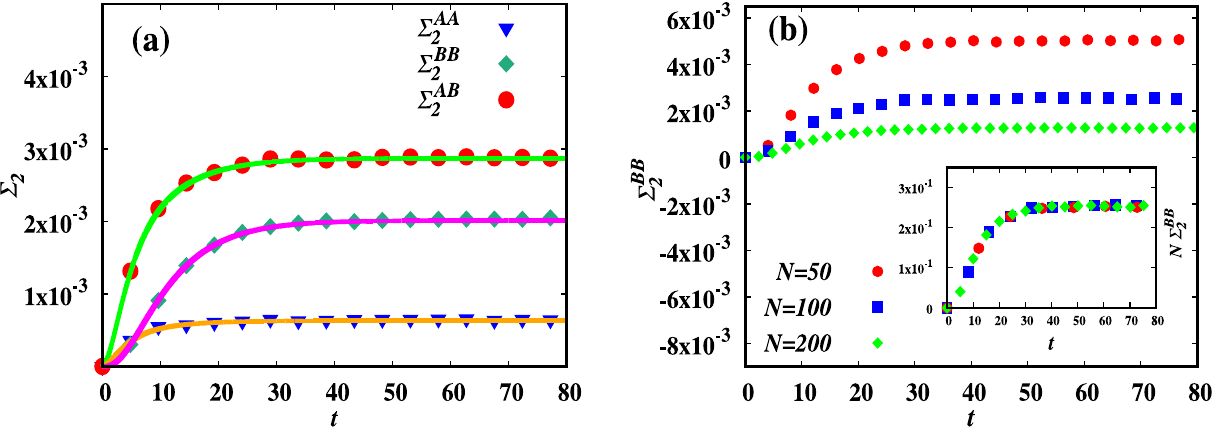} 
	\caption{\label{figure5}(a) The variation of the correlations $\Sigma^{AA}_2$, $\Sigma^{BB}_2$ and $\Sigma^{AB}_2$ with time $t$ for parameter values $r_{AB}=0.7$, $r_{AA}=r_{BB}=r_w=0.5$, $m_A=2$ and $m_B=1$. The noise distribution $\phi(\eta)$ is a normal distribution. The data points are from Monte Carlo simulations while the solid lines represent results from numerical integration of Eq.~(\ref{two-point-correlation-matrix-evolution}). (b) Monte Carlo data for the temporal dependence of $\Sigma^{BB}_2$ for three different $N$, the total number of particles. Inset:~The data for different $N$ collapse onto a single curve when scaled as in Eq.~(\ref{scaling}).
}
\end{figure}
Figure~\ref{figure6} shows the temporal dependence of  the average energies $\Sigma_1$ obtained from both Monte Carlo simulations as well as numerical integration of  Eq.~\eref{two-point-correlation-matrix-evolution} for two different sets of parameters. The numerically obtained steady state values are compared with the analytical results~[Eqs.~(\ref{66}) and (\ref{67})]. The steady state $\Sigma_1$ obtained from exact analysis is in agreement with Monte Carlo results. 
\begin{figure}
\centering
		\includegraphics[width=0.95 \columnwidth]{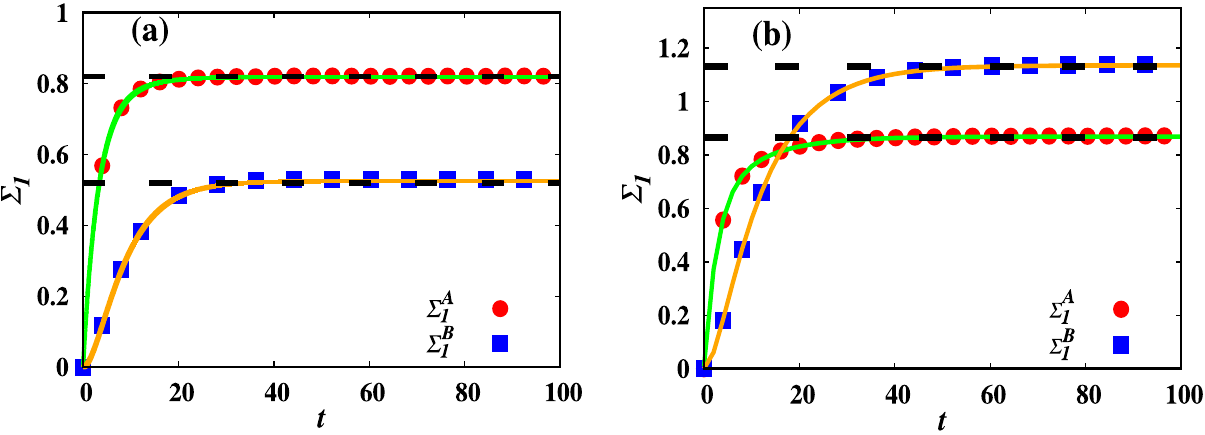} 
	\caption{\label{figure6}Variation of the mean energies $\Sigma^{A}_1$ and $\Sigma^{B}_1$ with time $t$. The data points are from Monte Carlo simulations, the solid lines represent results from numerical integration of Eq.~(\ref{two-point-correlation-matrix-evolution}), and the dashed lines denote the analytical results for the respective steady state values in Eqs.~(\ref{66}) and (\ref{67}). The noise distribution $\phi(\eta)$ is a normal distribution. 
(a) The data for parameter values $r_{AB}=0.4$, $r_{AA}=r_{BB}=r_w=0.5$, $m_A=2$ and $m_B=1$ illustrating the scenario when 
$\Sigma^A_1>\Sigma^B_1$.
(b) The data for parameter values $r_{AB}=0.7$, $r_{AA}=0.4$, $r_{BB}=1$, $r_w=0.5$, $m_A=2$ and $m_B=1$ for which 
Eq.~\eref{variance_condition} is satisfied, illustrating the scenario  when $\Sigma^B_1>\Sigma^A_1$.}
\end{figure}
We now show that though $B$ particles receive energy only through collisions with $A$, there is no order relation between the mean energies of $A$ and $B$ particles. From Eq.~(\ref{67}), it is easy to derive that $\Sigma^B_1\geq\Sigma^A_1$ whenever
\begin{equation}
X_A\geq \frac{1}{2}+\sqrt{\frac{1}{4} +\frac{\alpha_{BB} \lambda_{BB} \nu_B (1- \alpha_{BB})}{\lambda_{AB} \nu_A}}, \label{variance_condition}
\end{equation}
with $X_A$ as in Eq.~(\ref{eq: redefine x}). In Fig.~\ref{figure6}, we show results of simulations with choice of parameters such that $\Sigma^A_1>\Sigma^B_1$ [Fig.~\ref{figure6}(a)] or $\Sigma^B_1>\Sigma^A_1$ [Fig.~\ref{figure6}(b)], consistent with Eq.~(\ref{variance_condition}).

\section{\label{sec6-characteristic function} Analysis of the velocity distribution using characteristic function ($r_w=1$)}

When $r_w=1$, we show that the asymptotic behaviour of the velocity distributions can be determined analytically using characteristic functions. The characteristic functions for the velocity distributions are defined as
\begin{equation}
Z_k(q) \equiv \langle e^{-iqv}\rangle = \int^{+\infty}_{-\infty}P_k(v)e^{-iqv}dv,\quad k=A, B, \label{characteristic function}
\end{equation} 
where the subscript $k$ refers to either $A$ or $B$ type particles. Multiplying Eqs.~(\ref{pa_single}) by $e^{-iq v}$ and integrating over $v$, we obtain
\begin{align}
Z_k(q)=\frac{\nu_k\lambda_{kk}Z_k[(1-\alpha_{kk})q]Z_k(q\alpha_{kk})+\lambda_{k\bar{k}}\nu_{\bar{k}}Z_k[(1-X_{\bar{k}})q]Z_{\bar{k}}(qX_{\bar{k}})}{\nu_k\lambda_{kk}+\lambda_{k\bar{k}}\nu_{\bar{k}}+\delta_{kA}\lambda_d[1-f(q)]}, \label{eq:ZA}
\end{align}
where 
\be
f(q)\equiv \langle\exp(-iq\eta)\rangle_{\eta}
\ee
is the characteristic function for the noise distribution and $\bar{k}$ is as defined in Eq.~(\ref{eq:particle label}). The form for $f(q)$ depends on the parameter $\gamma$ [see Eq.~\eref{noise}]. 

Solving for $Z_A(q)$ and $Z_B(q)$ from Eq.~(\ref{eq:ZA}), we obtain
\begin{align}
&Z_A(q)=\frac{\nu_A\lambda_{AA}Z_A\big[\big(1-\alpha_{AA}\big)q\big]Z_A\big(\alpha_{AA}q\big)+\nu_B\lambda_{AB} Z_A\big[\big(1-\mu_B\alpha_{AB}\big)q\big]Z_B\big(\mu_B\alpha_{AB}q\big)}{\nu_A \lambda_{AA}+\nu_B \lambda_{AB}+\lambda_d\big(1- f(q)\big)}, \label{charac1}\\
&Z_B(q)=\frac{\nu_B \lambda_{BB}Z_B\big[\big(1-\alpha_{BB}\big)q\big]Z_B\big(\alpha_{BB}q\big)+\nu_A\lambda_{AB}Z_B\big[\big(1-\mu_A\alpha_{AB}\big)q\big]Z_A\big(\mu_A\alpha_{AB}q\big)}{ \nu_B\lambda_{BB}+\nu_A \lambda_{AB} }.   \label{charac2}
\end{align}
Equations~\eref{charac1} and \eref{charac2} are recursive in nature, and express the values
of the functions for a given value of $q$ in terms of smaller $q$. Since the value for small $q$
is known from the exact calculation of the second moment, the value of the characteristic function  
can in principle be calculated. However, in practice there are limitations in determining numerically the tails using this 
method~\cite{Prasad:17}.

The tails of the distribution can be obtained analytically by determining the singularities of the characteristic functions. On iterating  Eq.~\eref{charac1} with respect to $q$, it is clear that $Z_A(q)$ can be  expressed as an infinite product. Since the denominator in the right hand side of Eq.~\eref{charac1} involves a simple pole in the complex $q$ space, the terms in the infinite product representation of $Z_A(q)$ will correspond to a family of simple poles of which the one closest to the origin, $q_A^*$ determines the asymptotic behaviour of the velocity distribution. On the other hand, $Z_B$ does not have  any singular contribution from the denominator on the right hand side of Eq.~\eref{charac2}, and thus the leading singular behaviour of $Z_B$ is that of $Z_A(\alpha_{AB}\mu_Aq)$. This provides a relation between $q^*_A $ and the dominant pole of $Z_B(q)$, denoted as $q^*_B$, as
\begin{equation}
q^*_B=\frac{q^*_A}{\mu_A\alpha_{AB}}. \label{qb}
\end{equation}
The asymptotic tail of the distribution then decays exponentially as:
\begin{equation}
P_k(v)\sim e^{-|v|/v^*_k},~~ v^*_k=1/q^*_k, ~~k=A, B. \label{eq:prob dist}
\end{equation} 

The expression for $q^*_A$ depends on the specific values of the parameters, which we analyse now. If in Eq.~\eref{charac1}, 
if all the arguments of $Z_k$ on the right hand side are smaller than $q$, the dominant pole is given by equating the denominator to zero:
\begin{equation}
\nu_A\lambda_{AA}+\lambda_{AB}\nu_B+\lambda_d[1-f(q)]=0, ~~\alpha_{AA}<1.
\label{eq:lambda1}
\end{equation}
On the other hand when any of the arguments  of $Z_A$  on the right side of Eq.~\eref{charac1} is equal to $q$, which happens when $\alpha_{AA}=1$, then the equation satisfied by the pole is modified to (see  \ref{Calculation for the poles} for a derivation)
\begin{equation}
\nu_B\lambda_{AB}+\lambda_d[1- f(q)]=0, ~~\alpha_{AA}=1.
\label{pole:alpha=1}
\end{equation}
One may also find  a situation when the right hand side of Eq.~(\ref{charac1})  has an argument which is greater than $q$ for example when $1<\mu_k\alpha_{AB}<2$ [note that $\mu_k$ can take values between $(0,2)$]. However, one can show that the pole arising from these terms  happen to be further away from the pole obtained from Eq.~(\ref{eq:lambda1}), making the former ones irrelevant (see \ref{Calculation for the poles} for a detailed discussion). 

The above analysis presents two different kinds of behaviours which can be
realised in two different domains of the parameter space, one when $\alpha_{AA}<1$, and another when $\alpha_{AA}=1$. We present the details of the analysis in  \ref{Calculation for the poles}. While $q^*_k$ can be numerically determined for any $\gamma$, it takes on a simple form when $\gamma=1$ (exponential) or $\gamma=2$ (gaussian):
\begin{align}
  q^*_A = 
 \begin{cases}
c\sqrt{\frac{\nu_A \lambda_{AA} + \nu_B \lambda_{AB}}{\nu_A \lambda_{AA} + \nu_B \lambda_{AB} +\lambda_d}},&\alpha_{AA}<1,\gamma=1, \\ \label{lambda1}
\sqrt{-4c\ln\Big[{\frac{\lambda_d}{\nu_A \lambda_{AA} +\nu_B \lambda_{AB} + \lambda_d}}\Big]},&\alpha_{AA}<1,\gamma=2,  
\end{cases}
\end{align}
and
 \begin{align}
q^*_A =  
\begin{cases}
c\sqrt{\frac{ \nu_B \lambda_{AB}}{\nu_B \lambda_{AB} +\lambda_d}},&\alpha_{AA}=1,\gamma=1, \\ \label{eq: qa and qb 1}
\sqrt{-4c\ln\Big[{\frac{\lambda_d}{\nu_B \lambda_{AB} + \lambda_d}}\Big]},&\alpha_{AA}=1,\gamma=2. 
\end{cases}
\end{align}

The above analysis is valid only if all singularities of $f(q)$, the characteristic function of the noise distribution,  is larger in magnitude than $q_A^*$. This is clearly true for $\gamma>1$. For $\gamma=1$, the noise distribution is exponential and hence $f(q)$ has a simple pole at $c$, but it can be checked from Eqs.~(\ref{lambda1}) and (\ref{eq: qa and qb 1}) that $q_A^* <c$. When $\gamma<1$, $f(q)$ has a singular behaviour at $q=0$, and hence the singularity of $Z_A(q)$ and $Z_B(q)$ will be identical to that of $f(q)$. Hence, we conclude that if 
$\ln P_k(v) = - a_k |v|^\beta_k + \ldots$, then
\be
\beta_k=\min[\gamma,1], \quad r_w=1,~k=A, B.
\label{eq:betarw1}
\ee

For $\gamma\geq 1$, we check whether there is any order relation between $v^*_A$ and $v^*_B$. From Eq.~(\ref{qb}), it is clear that $v^*_A=v^*_B$ only for $\mu_A \alpha_{AB}=1$ or equivalently $r_{AB}=m_B/m_A$ [using Eqs.~(\ref{eq:alpha}) and (\ref{eq: redefine x})]. For other choices of parameters, both $v^*_A>v^*_B$ and $v^*_B>v^*_A$  can be realised, with $v^*_B>v^*_A$ for $r_{AB}>m_B/m_A$. 

The above analysis also implies  that the asymptotic behaviour of the velocity distribution of $A$ and $B$ type particles is independent of the coefficient of restitution $r_{BB}$ for $r_w=1$. Since $r_{BB}$ does not appear in Eqs.~(\ref{eq:lambda1}) and (\ref{pole:alpha=1}) for the poles, $q_A^*$ is independent of $r_{BB}$. Since the relation between $q_B^*$ and $q_A^*$ [see Eq.~(\ref{qb})] does not involve $r_{BB}$, $q_B^*$ is also independent of $r_{BB}$. Thus, for the tails of the distribution, one could have ignored all interactions among $B$ particles.

\section{Moment Analysis}\label{sec5-Moment Analysis}

The analysis based on characteristic functions in Sec.~\ref{sec6-characteristic function} works only for $r_w=1$. For $r_w\not=1$, we determine the tails of the distribution by analysing the asymptotic behaviour of large moments of velocity. The binary Maxwell gas
model, it turns out, allows for an exact numerical evaluation of the moments of the velocities. We outline the calculation below.

Let
\begin{equation}
M^k_{2n}\equiv \langle v^{2n}_k\rangle=\int^{+\infty}_{-\infty}dv~ v^{2n}P_k(v),\quad k=A,~B,
\end{equation}
be the $2n^{th}$ moment of the velocity distribution. The relation satisfied by the moments can be obtained by multiplying Eq.~(\ref{pa_single})  with $v^{2n}$ and integrating over all velocities.
 In the steady state, by setting the time derivative to zero, we obtain
\begin{align}
&\Big[\lambda_{kk} \nu_k  \Big(1-(1-\alpha_{kk})^{2n}-\alpha^{2n}_k \Big) + \lambda_{k\bar{k}}\nu_{\bar{k}} \Big(1-(1-X_{\bar{k}})^{2n}\Big)+ \delta_{kA}\lambda_d( 1-r^{2n}_w)\Big] \langle M^k_{2n}\rangle  \nonumber \\
& -\lambda_{k\bar{k}} \nu_{\bar{k}} X^{2n}_{\bar{k}} \langle M^{\bar{k}}_{2n}\rangle =\lambda_{kk}\nu_k \sum^{n-1}_{l=1}\binom {2n}{2l} (1-\alpha_{kk})^{2n-2l}\alpha^{2l}_k \langle M^k_{2n-2l}\rangle \langle M^k_{2l}\rangle      \nonumber \\
&+ \delta_{kA}\lambda_d \sum^{n-1}_{l=0}\binom {2n}{2l} r^{2l}_w \langle M^k_{2l}\rangle N_{2n-2l}+\lambda_{k\bar{k}} \nu_{\bar{k}} \sum^{n-1}_{l=1}\binom {2n}{2l} (1-X_{\bar{k}})^{2n-2l} X^{2l}_{\bar{k}} \langle M^k_{2n-2l}\rangle \langle M^{\bar{k}}_{2l}\rangle,\label{5.5a} 
\end{align}
where $\bar{k}$ is as defined in Eq.~\eref{eq:particle label}, and 
\begin{equation}
     N_{2n}=\langle  \eta^{2n}\rangle  = c^{\frac{-2n}{\gamma}}\frac{\Gamma(\frac{2n+1}{\gamma})}{ \Gamma(\gamma^{-1})}, \label{5.5ca}
\end{equation}
 is the $2n^{th}$ moment of the noise distribution. The moments thus satisfy a set of recurrence relations where $M_{2n}$ depends on moments of lower order. Clearly, $M^A_0=M^B_0=1$. Also, we know $M^A_2=\Sigma^A_1 $ and $M^B_2=\Sigma^B_1 $ from Eqs.~(\ref{66}) and (\ref{67}). Thus, by knowing these moments, we can generate all the higher moments.\par
The asymptotic behaviour of velocity distribution functions can be determined  by analysing the asymptotic behaviour of the ratios of large moments. 
We assume that the velocity distribution of $A$ and $B$ particles is asymptotically
a stretched exponential,~i.e.,
\begin{equation}
\ln P_k(v) = -a_k|v|^{\beta_k}+ \Psi_k(|v|),\quad a_k,~\beta_k>0,~k= A, B, \label{5.5pa}
\end{equation}
where the subleading correction $\Psi_k(|v|)$ satisfies $|v|^{-\beta_k}\Psi_k(|v|)\rightarrow0$ for large $v$. For such a stretched exponential distribution, the  $2n^{th}$ moment can be obtained using saddle point approximation for large $n$, and has the form~\cite{Prasad:19}
\begin{align}
        M^k_{2n}\sim \frac{n^{\frac{1}{\beta_k}}}{\sqrt{n}}\Big(\frac{2n}{a_k\beta_k}\Big)^{\frac{-2n}{\beta_k}}e^{\Psi\big[\big(\frac{2n}{a_k\beta_k}\big)^{1/\beta_k}\big]},\quad k=A,~B.
    \end{align}
    The ratio of consecutive moments is then given by
\begin{align}
       \Delta_{n,k}\equiv \frac{M^k_{2n}}{M^k_{2n-2}} = \Big( \frac{2n}{a_k\beta_k }\Big)^{2/\beta_k}\Big(1+\mathcal{O}(n^{-1})\Big),\quad k=A,~B. \label{5.5c}
\end{align}
It is clear from  Eq.~(\ref{5.5c}) that the ratio of large moments $\Delta_{n,k}$ depend only on 
the leading asymptotic behaviour of the velocity distribution and thus provides a tool for 
probing the tails of the distribution.

The parameters $\beta_k$ and $a_k$~[defined in Eq.~(\ref{5.5pa})] can be determined from $\Delta_{n,k}$~[Eq.~(\ref{5.5c})] for 2 consecutive
values of $n$ as
\begin{align}
\beta_k(n)&=\frac{ 2 \ln(n+1/n)}{\ln(\Delta_{n+1, k}/\Delta_{n, k})}, \label{exponent}\\
a_k(n)&=\frac{2n}{\beta_k (\Delta_{n, k})^{\beta_k/2}}. \label{prefactor}
\end{align}
The asymptotic $\beta_k$ and $a_k$ are obtained by extrapolating $\beta_k(n)$ and $a_k(n)$ to large $n$. We illustrate 
the procedure that we follow through an example. The dependence of parameters $\beta_k$ and $a_k$ on $n$ is shown in Figs.~\ref{figure20}(a) and \ref{figure20}(b) respectively for typical values of parameters. Clearly, they converge to their asymptotic value as $1/n$. Knowing the approach to the asymptotic value, the asymptotic value can be determined to high accuracy. We now use this approach to obtain the asymptotic behaviour of the velocity distributions for the cases of dissipative and diffusive driving.
\begin{figure}
\centering
		\includegraphics[width=0.95 \columnwidth]{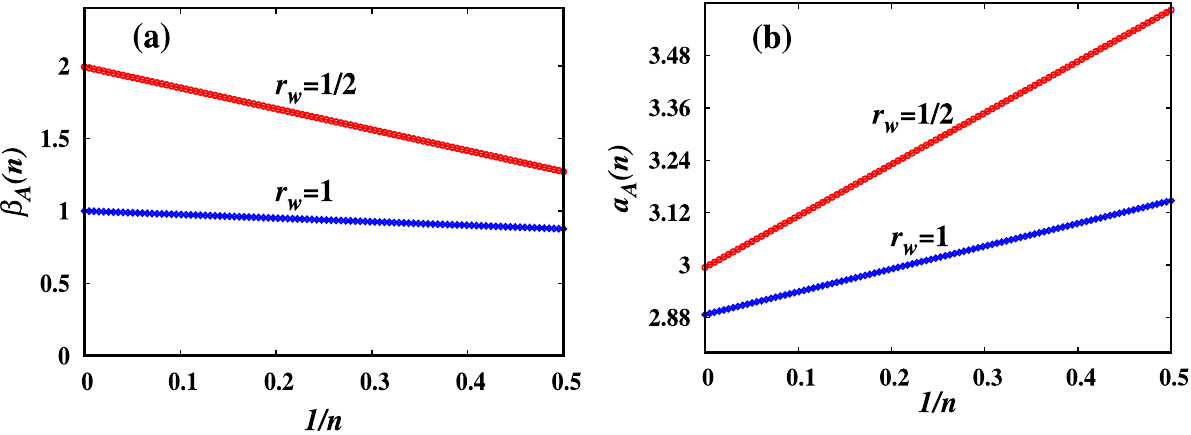}
	\caption{Variation of (a) the exponent $\beta_A(n)$ and (b) the prefactor $a_A(n)$, obtained numerically using Eqs.~\eref{exponent} and \eref{prefactor}, with $n^{-1}$  for the cases of dissipative ($r_w<1$) and diffusive ($r_w=1$) driving. The parameters for the noise distribution  was chosen to be $\gamma=2$ and $c=3$ [see Eq.~(\ref{noise})],  while the other parameters are $r_{AB}=r_{AA}=r_{BB}=0.5$, $m_A=2$ and $m_B=1$. $\beta_A(n)$ and $a_A(n)$ converge to their asymptotic value as $1/n$.}
\label{figure20}
\end{figure}

\subsection{Dissipative driving ($r_w <  1$)}

In this subsection, we discuss the results for dissipative driving when $r_w<1$. Figures~\ref{figure22} and \ref{figure24} show the results for $\beta_k$ and $a_k$ for $r_w=1/2$ and various choices of other parameters. In Fig.~\ref{figure22}(a), the dependence of $\beta_A$ and $\beta_B$ on $\gamma$, the exponent characterising the noise distribution [see Eq.~(\ref{exponent})] is shown. It is clear that $\beta_A=\beta_B=\gamma$, irrespective of whether $\gamma<1$ or $\gamma \geq 1$, i.e,
\be
\beta_k = \gamma, \quad r_w<1,~k=A, B.
\ee
We conclude that the velocity distribution is  non-universal.
\begin{figure}
\centering
		\includegraphics[width=0.95 \columnwidth]{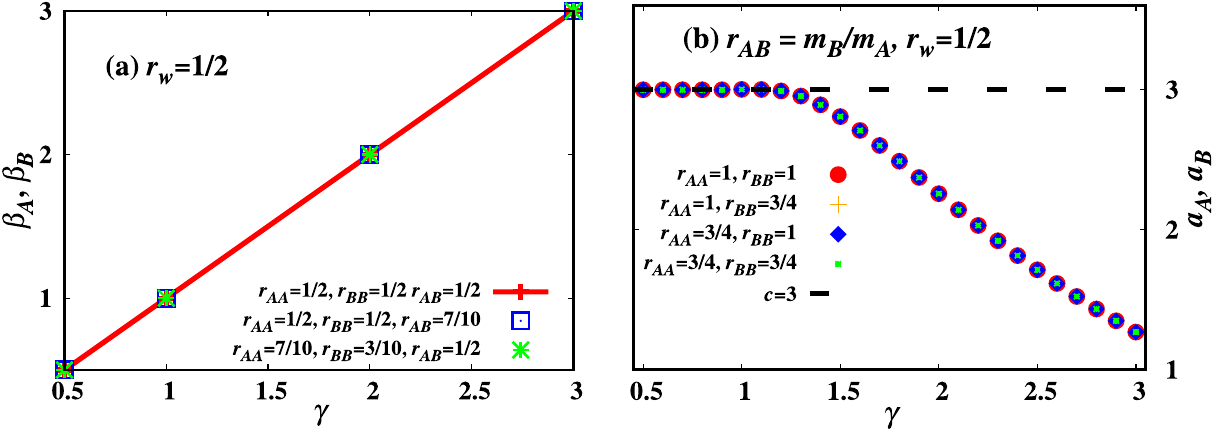}
	\caption{\label{figure22} Variation of (a) the exponents $\beta_k$ and (b) the prefactors $a_k$, characterising the asymptotic behaviour  of velocity distribution, with $\gamma$, the exponent characterising the noise distribution [see Eq.~(\ref{exponent})] for $r_w=1/2$, corresponding to dissipative driving.   The other common parameters are $c=3$,  $m_A=2$ and $m_B=1$. In (a), the data are for different choices of the coefficients of restitution $r_{AA}$, $r_{BB}$ and $r_{AB}$, while for (b) the data are for the special case $r_{AB}=m_B/m_A$. The dashed straight line corresponds to $c=3$.}
\end{figure}

In Sec.~\ref{sec6-characteristic function}, we showed that for $r_w=1$, the prefactors $a_A=a_B$ for the special case 
$r_{AB}=m_B/m_A$, and $a_A=a_B=c$ for $\gamma<1$. 
We check whether similar results hold for $r_w<1$. 
Figure~\ref{figure22}(b) shows the variation of the prefactors $a_A$ and $a_B$ with $\gamma$ 
for the special case $r_{AB}=m_B/m_A$, when $r_w=1/2$. 
Clearly, $a_A=a_B$ for all choices of $\gamma$. Also, for $\gamma<1$, $a_A=a_B$ as for diffusive driving. We, thus conclude that for the special case $r_{AB}=m_B/m_A$ and $r_w<1$, the asymptotic behaviour of the velocity distributions of both the components are identical.

We now determine the prefactor $a_k$ for other choices of parameters $r_{AB} \neq m_B/m_A$. Figure~\ref{figure24} shows the dependence of $a_A$ and $a_B$ on $\gamma$ for $r_w=1/2$ and different choices of other parameters. It is clear that $a_A>a_B$ for $r_{AB}>m_B/m_A$ [see Fig.~\ref{figure24}(a)]  and vice-versa [see Fig.~\ref{figure24}(b)].
\begin{figure}
\centering 
		\includegraphics[width=0.95 \columnwidth]{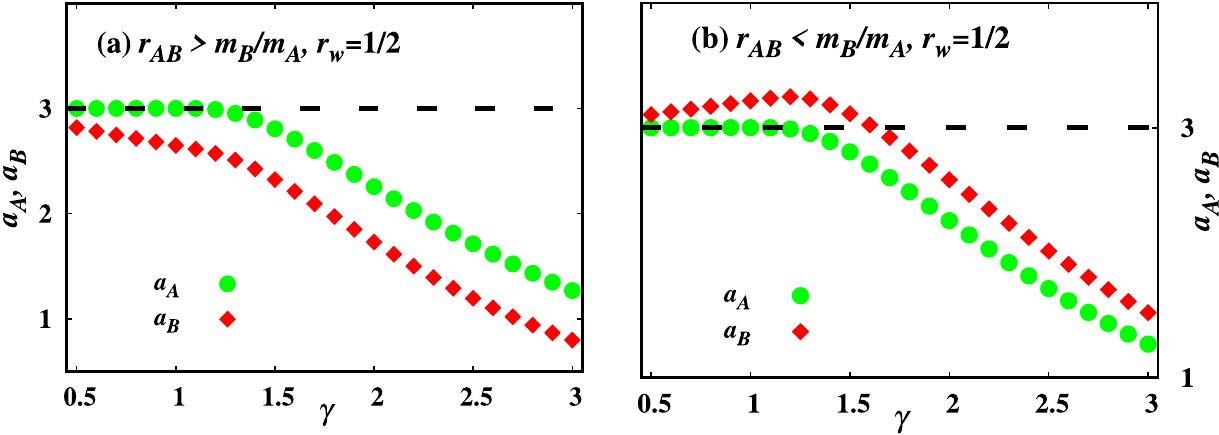}
	\caption{\label{figure24} Variation of the prefactor $a_k$ with $\gamma$, the exponent characterising the noise distribution [see Eq.~(\ref{prefactor})] for $r_w=1/2$, corresponding to dissipative driving.  The other common parameters are $c=3$,  $m_A=2$ and $m_B=1$. The dashed straight line corresponds to $c=3$. (a) Variation of $a_k$ with $\gamma$ for $r_{AB}=0.7$ such that $r_{AB}>m_B/m_A$. (b) Variation of $a_k$ with $\gamma$ for $r_{AB}=0.4$ such that $r_{AB}<m_B/m_A$.}
\end{figure}

Though we have presented results for only $r_w=1/2$, the results are similar for other choices of $r_w$. 
Thus, for $r_w<1$, both the components of the binary mixture have the same $\beta_k=\gamma$, the prefactors $a_k$ could be different and has no particular order relation. The velocity distribution is nonuniversal. 
 
\subsection{Diffusive driving ($r_w=1$)}

In this subsection, we discuss the results, using moment analysis, for diffusive driving when $r_w=1$. While we analysed this special case using characteristic functions, the analysis was not rigorous, and the numerical results will act as further confirmation. At the same time, the matching of the results obtained from both methods will act as a benchmark for the numerical moment-analysis method.

Figures~\ref{figure23} and \ref{figure25} show the results for $\beta_k$ and $a_k$ for $r_w=1$ and various choices of other parameters. In Fig.~\ref{figure23}(a), the dependence of $\beta_A$ and $\beta_B$ on $\gamma$, the exponent characterising the noise distribution [see Eq.~(\ref{exponent})] is shown. It is clear that $\beta_A=\beta_B=\min[\gamma,1]$, consistent with the analytical result in Eq.~(\ref{eq:betarw1}).
\begin{figure}
\centering
		\includegraphics[width=0.95 \columnwidth]{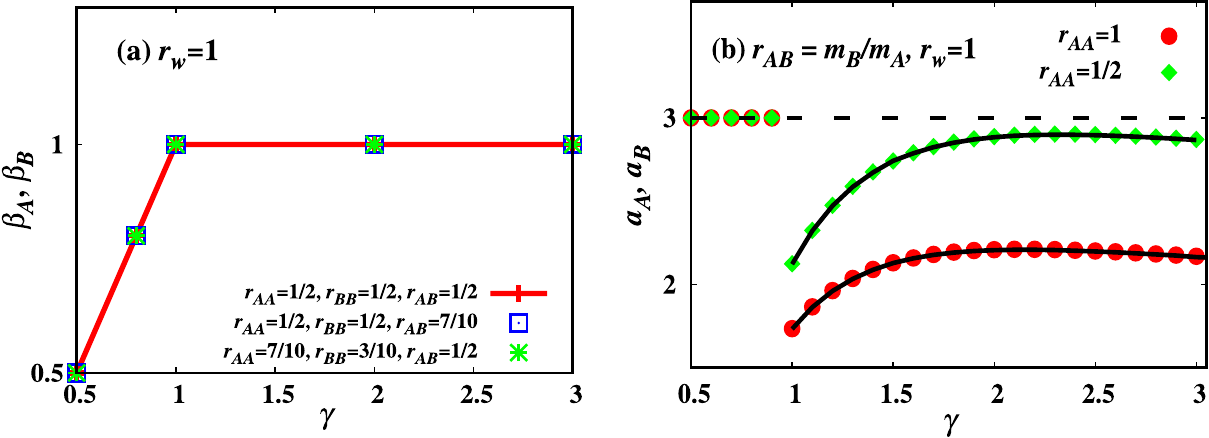}
	\caption{\label{figure23}Variation of (a) the exponents $\beta_k$ and (b) the prefactors $a_k$, characterising the asymptotic behaviour  of velocity distribution, with $\gamma$, the exponent characterising the noise distribution [see Eq.~(\ref{exponent})] for $r_w=1$, corresponding to diffusive driving.   The other common parameters are $c=3$,  $m_A=2$ and $m_B=1$. In (a), the data are for different choices of the coefficients of restitution $r_{AA}$, $r_{BB}$ and $r_{AB}$, while for (b) the data are for the special case $r_{AB}=m_B/m_A$. The dashed straight line corresponds to $c=3$. The data for $a_A$ and $a_B$ fall on top of each other for a given choice of $r_{AA}$ and $r_{AB}$. The solid lines in (b) are the exact solutions obtained by solving Eqs.~(\ref{eq:lambda1}) and (\ref{pole:alpha=1}).}
\end{figure}

In Sec.~\ref{sec6-characteristic function}, we showed that for $r_w=1$, the prefactors $a_A=a_B$ for the special case 
$r_{AB}=m_B/m_A$, and $a_A=a_B=c$ for $\gamma<1$. 
Figure~\ref{figure23}(b) shows the variation of the prefactors $a_A$ and $a_B$ with $\gamma$ 
for the special case $r_{AB}=m_B/m_A$, when $r_w=1$. 
Clearly, $a_A=a_B$ for all choices of $\gamma$. Also, for $\gamma<1$, $a_A=a_B=c$. In addition, the numerically obtained values for $a_A$ and $a_B$ coincide with that from the exact calculations (shown in solid lines in Fig.~\ref{figure23}(b)).

We now determine the prefactor $a_k$ for other choices of parameters $r_{AB} \neq m_B/m_A$. Figure~\ref{figure25} shows the dependence of $a_A$ and $a_B$ on $\gamma$ for $r_w=1$ and different choices of other parameters. It is clear that $a_A>a_B$ for $r_{AB}>m_B/m_A$ [see Fig.~\ref{figure25}(a)]  and vice-versa [see Fig.~\ref{figure25}(b)]. Also, the numerically obtained values coincide with the analytical results (shown in solid lines in Fig.~\ref{figure25})
\begin{figure}
\centering 
		\includegraphics[width=0.95 \columnwidth]{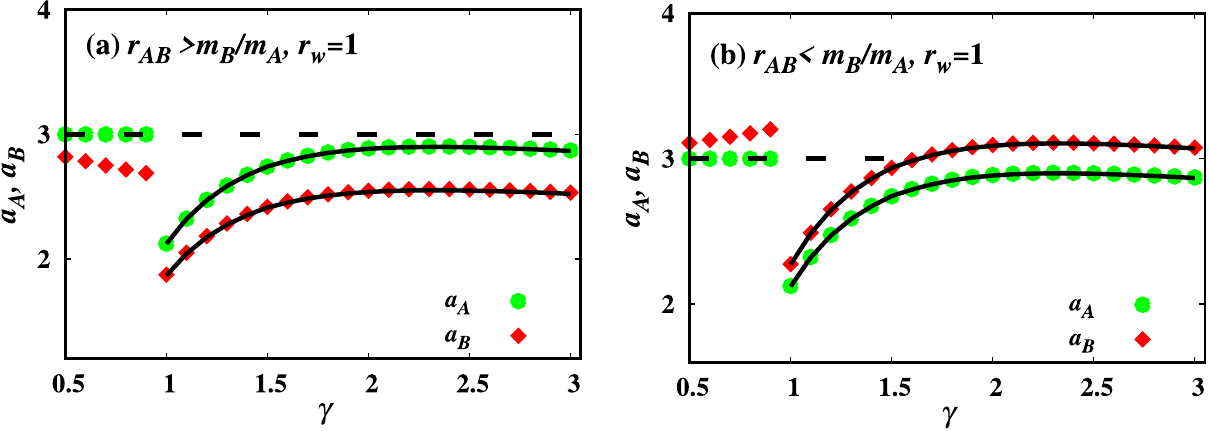} 
\caption{\label{figure25} Variation of the prefactor $a_k$ with $\gamma$, the exponent characterising the noise distribution [see Eq.~(\ref{prefactor})] for $r_w=1$, corresponding to diffusive driving.  The other common parameters are $c=3$,  $m_A=2$ and $m_B=1$. The solid lines  are the exact solutions obtained by solving Eq.~(\ref{eq:lambda1}) and the dashed straight line corresponds to $c=3$. (a) Variation of $a_k$ with $\gamma$  for $r_{AB}=0.7$ such that $r_{AB}>m_B/m_A$. (b) Variation of $a_k$ with $\gamma$ for $r_{AB}=0.4$ such that $r_{AB}<m_B/m_A$.}
\end{figure}

Thus, for diffusive driving, both the components of the binary mixture show similar asymptotic behaviour for the parameter $r_{AB}=m_B/m_A$ as $\beta_A=\beta_B$ and $a_A=a_B$.  For other choices of the parameter $r_{AB}$, $\beta_A=\beta_B$ but $a_A\neq a_B$. So, in that case, velocity distribution of one of the components of the binary mixture decay slower in comparison to the other. Moreover, the asymptotic velocity distribution is  independent on the noise distribution as $\beta_A=\beta_B=1$ provided the noise distribution $\phi(\eta)$ decays faster than exponential and hence the velocity distribution is universal.

\section{\label{sec7-Analysis of tail by truncating the driving term}Analysis of tail of distribution for diffusive driving}

In this section, we derive the asymptotic behaviour of the velocity distributions when the driving is modelled by a phenomenological diffusive term, $D \frac{\partial^2P}{\partial v^2}$ as in Eq.~(\ref{driving3}), as is  customarily done in kinetic theory of dilute inelastic gases.   We use this form of driving in Eq.~(\ref{pa_single}) to analyse the tail of the velocity distributions. Consider first $P_A(v)$, the velocity distribution of $A$ particles. Its time evolution is as given in Eq.~(\ref{pa_single}) with $k=A$. It can be shown that the gain terms arising from collisions with other particles are sub-dominant  compared to the corresponding loss terms~\cite{Noije:98}.  Dropping these terms, in the steady state, Eq.~(\ref{pa_single}) takes the form
\begin{equation}
 0 \approx -\left(\lambda_{AA} \nu_A +   \lambda_{AB} \nu_B \right)  P_A(v) + D \frac{d^2}{dv^2}P_A (v). \label{pa1-single}
\end{equation}
By making the ansatz $P_A(v)\sim\exp(-a|v|^\alpha)$ for large $|v|$ in Eq.~(\ref{pa1-single}), we obtain $\alpha=1$ (see Ref.~\cite{Noije:98} for more details), or equivalently $P_A(v)\sim \exp(-a|v|)$. 

For $B$ type particles, there is no diffusive term arising from driving as $B$ particles are not driven externally. Therefore, we cannot drop the gain term arising from collision with $A$ type particles in Eq.~(\ref{pa_single}). Dropping the sub-dominant gain term arising from $B$ type particles, Eq.~(\ref{pa_single}) takes the form
 \begin{equation}
 0 \approx -\left( \lambda_{BB} \nu_B + \lambda_{AB}\nu_A \right) P_B(v)+\lambda_{AB} \nu_A \int_{-\infty}^{\infty} dv_1 P_A(v_1) P_B\Big(\frac{v-X_Av_1}{1-X_A}\Big). \label{pb1-single}
 \end{equation}
We assume that $P_B(v)$ has the asymptotic form $P_B (v) \sim \exp(-b|v|^\beta)$. Using $P_A(v)\sim \exp(-a|v|)$, Eq.~(\ref{pb1-single}) simplifies to
 \begin{equation}
P_B(v) \sim  \int dv_1 e^{-a |v_1|}e^{-b\left |\frac{v-X_Av_1}{1-X_A}\right |^\beta}. \label{pb2-single}
 \end{equation}
Substituting $v_1=x v$, $v\gg1$, we obtain
 \begin{equation}
e^{-b v^\beta} \sim  \int dx e^{-a |x| v -\frac{b v^\beta |1-X_A x|^\beta}{|1-X_A|^\beta}}. \label{pb3-single}
 \end{equation}

We now analyse Eq.~(\ref{pb3-single}) for the cases $\beta<1$, $\beta>1$ and $\beta=1$. When $\beta<1$, the integrand on the right hand side of Eq.~(\ref{pb3-single}) is maximised at $x=0$. Then we obtain $\exp[-b v^\beta]\sim  \exp[-b v^\beta/|1-X_A|^\beta]$. A self consistent solution is possible only if $X_A=0$, but from Eq.~(\ref{eq: redefine x}) we know that $X_A>0$. Hence, there exists  no solution for $\beta<1$. Now, consider $\beta>1$. In this case the right hand side of Eq.~(\ref{pb3-single}) is maximised at $x=1/X_A$. Then we obtain $\exp[-b v^\beta]\sim  \exp[-a x v]$, giving $\beta=1$, in contradiction to our assumption that $\beta>1$. Hence, there exists  no solution for $\beta > 1$. For $\beta=1$, it is easy to see the integrand on the right hand side of  Eq.~(\ref{pb3-single}) can be evaluated by a saddle point integration. This will lead to a self-consistent equation obeyed by $b$. Thus, we conclude that $\beta=1$ for $B$ particles also. 

To summarise, if the driving is modelled by a diffusive term, then kinetic theory predicts that the asymptotic behaviour of the velocity distribution of both $A$ and $B$ particles are universal and exponential.

\section{\label{sec8-Discussions}Summary and discussion} 

In this paper, we studied an inelastic driven one dimensional Maxwell gas consisting of two components. Only one of the two components is driven externally, while the other component receives energy through inter-particle collisions. The well-mixed limit is assumed such that spatial correlations were ignored. The main aim of the paper was to determine the steady state velocity distribution $P(v)$ for both the components. The behaviour for small velocities is captured by lower order moments of the velocity. The second moment of the velocity, as well as the different two point correlations, were determined exactly. This was possible because the equations for the two point correlations form a closed set. We showed that for suitable choice of parameters, the mean granular temperature of either component could be larger, even though only one component is driven. We also showed that two point correlations involving different particle types vanish in the thermodynamic limit.

The asymptotic behaviour of the tails of $P(v)$, characterised by  $\ln P(v)=  -a |v|^\beta + \ldots$ for large $|v|$, was determined either exactly for certain ranges of parameters, or through numerical analysis of the ratio of consecutive large moments of the velocity. The latter was possible because a given moment can be calculated, to any desired accuracy, if all moments of lower order are known, allowing a recursive calculation to be implemented. The results depend on the details of driving, which we had implemented as follows: when a particle with velocity $v$ is driven, its velocity is modified to $-r_w v+ \eta$, where the noise $\eta$ is chosen from a stretched exponential distribution $\phi(\eta)$ that has the asymptotic behaviour $\ln \phi(\eta) \propto -|\eta|^\gamma$ for large $\eta$.  For diffusive driving ($r_w=1$), we showed, using characteristic functions, that $\beta=\min[1,\gamma]$ for both components, implying that the tails are universal for $\gamma>1$.  For inelastic driving ($r_w<1$), we show that $\beta=\gamma$ for both components, implying that the distributions are always non-universal. The constant $a$ is different for the two components in both cases of driving, and depending on the values of coefficients of restitution, could be larger for either component. We show that only when $r_{AB}=m_B/m_A$, then the constant $a$ is same for both components, and the asymptotic behaviour of both components become identical. One also observes that the distribution for $A$ and $B$ particles for the case of diffusive driving ($r_w=1$) are independent of the parameter $r_{BB}$. It is because the equations characterising the poles for $A$ and $B$ particles are independent of the parameter $r_{BB}$. This observation indicates that the steady state velocity distribution is independent of the interparticle collisions amongst the $B$ particles.

The results that we have obtained are compared with the results from Boltzmann equation where the driving term is modelled by a diffusive term, as is usually done in kinetic theory. The Boltzmann equation predicts that the velocity distribution for both components should have an exponential distribution, irrespective of noise distribution. These results coincide with the detailed results for our model only for the case $\gamma\geq1$ and $r_w=1$. Thus, the truncation of the driving term in the Boltzmann equation to lowest order in $\eta$ gives the correct result only in restricted regimes. However, even this restricted equivalence between microscopic models for driving and Boltzmann equation with diffusive driving may not hold for more realistic collision kernels where the collision rates are proportional to the relative velocity~\cite{Prasad:18,Prasad:19}.

In experiments on bilayers, where only the bottom layer is driven, it has been observed that the data for the velocity distribution for the bottom layer are consistent with  $\beta \approx 1.5$, while that for the top layer is consistent with $\beta \approx 2$. These are in contradiction to the fact that in the Maxwell model studied in this paper, $\beta$ for the velocity distributions of  both components are the same. However, one can also consider a particular limit in which the driving rate $\lambda_d$ becomes large when compared to the collision rates, where  the driving may be generically described as an Ornstein Uhlenbeck process for any noise distribution $\phi(\eta)$ with finite second moment (see \ref{Large driving limit} for detailed analysis). For a binary gas with both components driven by an Ornstein Uhlenbeck process, the velocity distributions for both the components has been shown to follow a gaussian statistics~\cite{Marconi:02}. We find that the velocity distribution remains Gaussian for both the components even when only one of the components are driven as in bilayer system. We also note that it is highly unlikely that the velocity distribution of $B$ particles will decay with a larger $\beta$, i.e. $\beta_B > \beta_A$. This is because there is always a contribution to the tails coming from $B$ particles that have just undergone a velocity transferring collision with an $A$ particle.

In the Maxwell model considered in the paper, we have assumed that the collision rate of a pair of particles is independent of the relative velocity. For ballistic transport, the collision rate is proportional to the relative velocity. For mono dispersed gases, an analysis with this more realistic kernel shows that $\beta$ remains the same, though for $r_w=1$, there are additional logarithmic corrections to the exponential decay~\cite{Prasad:18,Prasad:19}. We expect these results to generalise to the driven binary gas also, such that $\beta$, as obtained in this paper, is not modified. Showing this more rigorous is a promising area for future study.

A significant simplification is restricting the model to one dimension. In this case, all collisions are head on. For mono-dispersed gases, the possibility of glancing collisions in two and higher dimensions introduces another universal regime for the velocity distribution which is a Gaussian distribution with logarithmic corrections~\cite{Prasad:18}. We thus, expect that the velocity distributions for the binary gas also becomes near gaussian in two and higher dimensions.

The spatial correlations that have been ignored in our calculations, can be studied only through large scale simulations. However, conventional simulations sample only the typical velocities making it difficult to sample the tails. Biased simulations which give extra weight to rare events might overcome this difficulty. We are currently working on such numerical approaches.


\begin{appendices}
\appendix
  
\section{\label{Calculation for the poles}Analysis of the tails of the velocity distribution  when $r_w=1$}

In this appendix, we discuss in detail how to obtain the tails of velocity distribution for the case of $r_w=1$,  as given in Eqs.~(\ref{eq:prob dist}) and (\ref{eq: qa and qb 1}-\ref{qb}), using characteristic functions. For $r_w=1$ the characteristic function of the velocity distribution $Z_k(q)$ [see Eq.~\eref{characteristic function}] with $k=A,B$ satisfy
\begin{align}
&Z_A(q)=\frac{\nu_A\lambda_{AA}Z_A\big[\big(1-\alpha_{AA}\big)q\big]Z_A\big(\alpha_{AA}q\big)+\nu_B\lambda_{AB} Z_A\big[\big(1-\mu_B\alpha_{AB}\big)q\big]Z_B\big(\mu_B\alpha_{AB}q\big)}{\nu_A \lambda_{AA}+\nu_B \lambda_{AB}+\lambda_d\big(1- f(q)\big)}, \label{ap-charac1}\\
&Z_B(q)=\frac{\nu_B \lambda_{BB}Z_B\big[\big(1-\alpha_{BB}\big)q\big]Z_B\big(\alpha_{BB}q\big)+\nu_A\lambda_{AB}Z_B\big[\big(1-\mu_A\alpha_{AB}\big)q\big]Z_A\big(\mu_A\alpha_{AB}q\big)}{ \nu_B\lambda_{BB}+\nu_A \lambda_{AB} },   \label{ap-charac2}
\end{align}
which are the same as in Eqs.~\eref{charac1} and \eref{charac2} respectively.
As discussed in \sref{sec6-characteristic function},  the tails of the velocity distribution are determined by the pole in $Z_k(q)$ closest to the origin defined as  $q^*_k$. Then,  $P_k(v)\sim e^{-q^*_k |v|}$ for large $|v|$. 
In what follows we first consider the case 
$\alpha_{AA}<1$ and then $\alpha_{AA}=1$ to evaluate $q^*_k$ for each case.

\subsection{$\boldsymbol{\alpha_{AA}<1}$}

In this case, given $\mu_B\alpha_{AB}<1$, all the arguments of  $Z_A$ and $Z_B$ on the right side of Eq.~(\ref{ap-charac1}) are less than $q$ and hence the pole closest to 
 the origin for $Z_A$, i.e., $q^*_A$ is given by equating its denominator to zero as 
\begin{equation}
\nu_A\lambda_{AA}+\lambda_{AB}\nu_B+\lambda_d[1-f(q_A^*)]=0.
\label{appendix:lambda1}
\end{equation}  
All other poles are larger in magnitude than $q_A^*$ that satisfies Eq.~(\ref{appendix:lambda1}). 

For general $f(q_A^*)$, Eq.~(\ref{appendix:lambda1}) has to be solved numerically. But, when the noise distribution is a gaussian ($\gamma=2$) or exponential ($\gamma=1$), then $f(q)$ is $e^{-q^2/4c}$ and $c^2/(c^2 + q^2)$ respectively. In this case $q_A^*$ has a simple form:
\begin{align}
q^*_A = 
\begin{cases} 
c\sqrt{\frac{\nu_A \lambda_{AA} + \nu_B \lambda_{AB}}{\nu_A \lambda_{AA} + \nu_B\lambda_{AB} +\lambda_d}},
&\gamma=1, \\ \label{ap-lambda1}
\sqrt{-4c\ln\Big[{\frac{\lambda_d}{\nu_A \lambda_{AA} +\nu_B \lambda_{AB} + \lambda_d}}\Big]},
&\gamma=2,
\end{cases}
\end{align}
as displayed in Eq.~\eref{lambda1}.

Now, we look for the pole of $Z_B$ in Eq.~(\ref{ap-charac2}). The expression for $Z_B$ in Eq.~(\ref{ap-charac2}) itself does not have any singularity but the pole arises from the dependence of $Z_B(q)$ on $Z_A(\alpha_{AB}\mu_Aq)$ which is obtained using Eq.~(\ref{ap-charac1}) as
\begin{align}
&\left[\nu_A\lambda_{AA}+\lambda_{AB}\nu_B+ \lambda_d[1-f(\alpha_{AB}\mu_Aq)]\right] Z_A(\mu_A \alpha_{AB}q)=
\nu_A\lambda_{AA}Z_A\Big[\big(1-\alpha_{AA}\big)\alpha_{AB}\mu_Aq)\Big]\nonumber \\
& \times Z_A\Big(\alpha_{AA}\alpha_{AB}\mu_Aq \Big) 
+\lambda_{AB}\nu_BZ_A\Big[\big(1-\mu_B\alpha_{AB}\big)\alpha_{AB}\mu_Aq  \Big] Z_B\Big(\alpha^2_{AB}\mu_A\mu_Bq \Big).
\label{appendix:charac5}
\end{align}
As all the arguments of $Z_A$ and $Z_B$ on the right side of Eq.~(\ref{appendix:charac5}) are less than the argument of $Z_A$ on the left side, 
the poles originating from the latter terms will be further away from the origin compared to the one obtained by equating the denominator of Eq.~(\ref{appendix:charac5}) to zero. This results in the relation satisfied by $q^*_B$ as:
 \begin{equation}
 \nu_A\lambda_{AA}+\lambda_{AB}\nu_B+\lambda_d[1-f(\alpha_{AB}\mu_Aq)]=0,\quad q=q^*_B,
 \label{appendix:qb}
 \end{equation} 
and using Eq.~\eref{ap-lambda1},  $q^*_B$ follows the form:
\begin{equation}
q^*_B=\frac{q^*_A}{\mu_A\alpha_{AB}}. \label{ap-qb}
\end{equation}
 
However, for $1<\mu_k\alpha_{AB}<2$, the argument of $Z_B$ on the right hand side of Eq.~(\ref{ap-charac1}) becomes larger than $q$ and the pole closest to the origin may arise from this term.  We compute $Z_B(\mu_B \alpha_{AB}q)$ from Eq.~\eref{ap-charac2},  and has the form
\begin{align}
(\nu_A\lambda_{AB} + \nu_B\lambda_{BB}) Z_B(\mu_B \alpha_{AB}q)&=
  \nu_B\lambda_{BB}Z_B\big[\big(1-\alpha_{BB}\big)\mu_B\alpha_{AB}q\big]Z_B\big(\alpha_{BB}\alpha_{AB}\mu_Bq\big) \nonumber \\
  &+\nu_A\lambda_{AB}Z_B\big[\big(1-\alpha_{AB}\mu_A\big)\mu_B\alpha_{AB}q\big]Z_A\big(\alpha^2_{AB}\mu_A\mu_Bq\big).
 \label{appendix:charac4}
 \end{align}
Equation~(\ref{appendix:charac4}) shows that the pole arises from the further dependence of $Z_B$ on $Z_A$ [see Eq.~(\ref{ap-charac1})]. Since, the argument of $Z_A$, i.e., $\alpha^2_{AB}\mu_A\mu_Bq<q$, the pole originating from the term $Z_B(\mu_B \alpha_{AB}q)$ is further away from the origin in comparison to the one obtained by equating the denominator of Eq.~(\ref{ap-charac1}) to zero. Hence, the closest pole to the origin for $Z_A$, i.e., $q^*_A$ follows Eq.~\eref{ap-lambda1}.

Now, the poles of $Z_B$~[Eq.~\eref{ap-charac2}] for $1<\mu_k\alpha_{AB}<2$ arises from its dependence on $Z_A(\alpha_{AB}\mu_Aq)$ 
[Eq.~(\ref{appendix:charac5})].  Here, it can be shown that the terms $\alpha_{AA}$, $(\mu_A\alpha_{AB}-1)$ and $\alpha^2_{AB}\mu_A\mu_B$ present in the arguments of $Z_A$ and $Z_B$ on the right side of Eq.~(\ref{appendix:charac5}) are less than unity and hence make the arguments of $Z_A$ smaller on the right side when compared to that on the left side. Thus the poles originating from the latter terms present will be further away from the origin compared to the one obtained by equating the denominator of Eq.~(\ref{appendix:charac5}) to zero. Thus, the pole of $Z_B$ for this case follows Eq.~\eref{ap-qb}.

For the special case of $\alpha_{BB}=1$, Eq.~\eref{ap-charac2} takes the form \begin{equation}
(\nu_A\lambda_{AB})Z_B(q)=\nu_A\lambda_{AB}Z_B\big[\big(1-\alpha_{AB}\mu_A\big)q\big]Z_A\big(\alpha_{AB}\mu_Aq\big), \label{appendix:charac7}
\end{equation}
whereas, Eq.~(\ref{ap-charac1}) remains the same. Since the pole of $Z_B$ appears from its dependence on $Z_A$ and the argument of $Z_A$ in Eq.~(\ref{appendix:charac7}) is the same as in Eq.~(\ref{ap-charac2}),  this case is similar to the one analysed by considering generic range of $\alpha_{BB}$. Thus, the poles in this case are also obtained by solving Eqs.~(\ref{appendix:lambda1}) and (\ref{appendix:qb}).

\subsection{$\boldsymbol{\alpha_{AA}=1}$}

Next we consider the case $\alpha_{AA}=1$. For this case, Eq.~(\ref{ap-charac1}) takes the form
\begin{equation}
Z_A(q)=\frac{\lambda_{AB}\nu_B Z_A\big[\big(1-\mu_B\alpha_{AB}\big)q\big]Z_B\big(\mu_B\alpha_{AB}q\big)}{\nu_B\lambda_{AB}+\lambda_d\big(1- f(q)\big)},  \label{appendix:charac6}
\end{equation}
and the expression for $Z_B$ remains the same as given in Eq.~(\ref{ap-charac2}). Here, we follow the same reasoning as for the case of $\alpha_{AA}<1$ to find the closest poles to the origin as the arguments of $Z_j's$, $j=A,~B$ on the right side of Eqs.~(\ref{ap-charac2}) and (\ref{appendix:charac6}) are still the same. Now the pole is different from the previous case $\alpha_{AA}<1$, as the source of singularity, i.e., the denominator of Eq.~(\ref{appendix:charac6}), is now modified. The poles are obtained by equating the denominator of Eq.~(\ref{appendix:charac6}) to zero, i.e., 
\begin{equation}
\nu_B\lambda_{AB}+\lambda_d[1- f(q)]=0, \label{appendix:alpha=1}
\end{equation}
and results in the form for $q^*_A$:
\begin{align}
q^*_A =  
\begin{cases}
c\sqrt{\frac{ \nu_B \lambda_{AB}}{\nu_B \lambda_{AB} +\lambda_d}},&\gamma=1, \\ \label{ap-eq: qa and qb 1}
\sqrt{-4c\ln\Big[{\frac{\lambda_d}{\nu_B \lambda_{AB} + \lambda_d}}\Big]},&\gamma=2,
\end{cases}
\end{align}
with $q_B$ given as in Eq.~\eref{ap-qb}.

For the special case of $\alpha_{BB}=1$, Eq.~\eref{ap-charac2} takes the form as given in Eq.~(\ref{appendix:charac7}) whereas Eq.~(\ref{appendix:charac6}) remains the same. One can follow the same analysis as described for generic range of $\alpha_{BB}$ to show that the poles in this case are also obtained by solving Eq.~(\ref{appendix:alpha=1}).

\section{\label{Large driving limit}Limiting case of large driving}

In this appendix, we examine the limit when the driving rate is large.  For the driving studied in the paper
[see Eq.~\eref{driving}], for noise distributions with finite second moment $\sigma^2$,
one can show that in the limit with $\lambda_d \to \infty$, $\sigma^2\to 0$, and $r_w\to-1$, keeping $\lambda_d[(1+r_w)]= \Gamma$ and $\lambda_d\sigma^2/2 = D$, the driving approaches an Ornstein-Uhlenbeck driving
where the 
change in velocity of the particles due to driving is effectively described by
$\partial v/\partial t=\Gamma v +\xi$, where $\xi$ is an effective noise with $\langle\xi^2\rangle=2D$~\cite{Prasad:14}.
The tails of the velocity distributions for a binary Maxwell gas with both the components driven by Ornstein-Uhlenbeck  driving has been shown to be a  Gaussian~\cite{Marconi:02}. In the following, we present an   analysis of a binary Maxwell gas with Ornstein-Uhlenbeck driving that acts only on  one  species of particles, and show that  the velocity distribution for both the particles have
gaussian tails.

We consider the limit of large driving in Eq.~(\ref{pa_single}) to analyse the tail of the velocity distributions. Consider first $P_A(v)$, the velocity distribution of $A$ particles. Its time evolution is as given in Eq.~(\ref{pa_single}) with $k=A$.  As discussed in \sref{sec7-Analysis of tail by truncating the driving term}, it  can be shown that 
the gain terms in Eq.~(\ref{pa_single}) arising from collisions with other particles are sub-dominant  compared to the corresponding loss terms~\cite{Noije:98}.  Dropping these terms and taking the limit of large driving, in the steady state, Eq.~(\ref{pa_single}) takes the form
\begin{equation}
 0 \approx -\left(\lambda_{AA} \nu_A +   \lambda_{AB} \nu_B \right)  P_A(v) + D \frac{d^2}{dv^2}P_A (v) + \Gamma\frac{d}{dv}[vP_A(v)]. \label{ap-A}
\end{equation}
By making the ansatz $P_A(v)\sim\exp(-a|v|^\alpha)$ for large $|v|$ in Eq.~(\ref{ap-A}), we obtain $\alpha=2$ (see Ref.~\cite{Noije:98} for more details), or equivalently $P_A(v)\sim \exp(-av^2)$. 

For $B$ type particles, there is no diffusive term arising from external driving, as $B$ particles are not driven externally. Therefore, 
we cannot ignore  the gain term arising from collision with $A$ type particles in Eq.~(\ref{pa_single}). Dropping the sub-dominant gain term arising from $B$ type particles, Eq.~(\ref{pa_single}) takes the form
 \begin{equation}
 0 \approx -\left( \lambda_{BB} \nu_B + \lambda_{AB}\nu_A \right) P_B(v)+\lambda_{AB} \nu_A \int_{-\infty}^{\infty} dv_1 P_A(v_1) P_B\Big(\frac{v-X_Av_1}{1-X_A}\Big). \label{ap-B1}
 \end{equation}
We assume that $P_B(v)$ has the asymptotic form $P_B (v) \sim \exp(-b|v|^\beta)$. Using $P_A(v)\sim \exp(-av^2)$, Eq.~(\ref{ap-B1}) simplifies to
 \begin{equation}
P_B(v) \sim  \int dv_1 e^{-a v_1^2}e^{-b\left |\frac{v-X_Av_1}{1-X_A}\right |^\beta}. \label{ap-B}
 \end{equation}
Substituting $v_1=x v$, $v\gg1$, we obtain
 \begin{equation}
e^{-b v^\beta} \sim  \int dx e^{-a x^2 v^2 -\frac{b v^\beta |1-X_A x|^\beta}{|1-X_A|^\beta}}. \label{ap-B2}
 \end{equation}

We analyse Eq.~(\ref{ap-B2}) for the cases $\beta<2$, $\beta>2$ and $\beta=2$. When $\beta<2$, the integrand on the right hand side of Eq.~(\ref{ap-B2}) is maximised at $x=0$. Then we obtain $\exp[-b v^\beta]\sim  \exp[-b v^\beta/|1-X_A|^\beta]$. A self consistent solution is possible only if $X_A=0$, but from Eq.~(\ref{eq: redefine x}) we know that $X_A>0$. Hence, there exists  no solution for $\beta<2$. Now, consider $\beta>2$. In this case the right hand side of Eq.~(\ref{ap-B2}) is maximised at $x=1/X_A$. Then we obtain $\exp[-b v^\beta]\sim  \exp[-a x^2 v^2]$, giving $\beta=2$, in contradiction to our assumption that $\beta>2$. Hence, there exists  no solution for $\beta > 2$. For $\beta=2$, it is easy to see the integrand on the right hand side of  Eq.~(\ref{ap-B2}) can be evaluated by a saddle point integration. This will lead to a self-consistent equation obeyed by $b$. Thus, we conclude that $\beta=2$ for $B$ particles or equivalently $P_B(v)\sim\exp(-bv^2)$. Therefore, the tails of the velocity distributions of both the components are gaussian.

\end{appendices}

\bibliographystyle{iopart-num}

\begin{thebibliography}{10}
\expandafter\ifx\csname url\endcsname\relax
  \def\url#1{{\tt #1}}\fi
\expandafter\ifx\csname urlprefix\endcsname\relax\def\urlprefix{URL }\fi
\providecommand{\eprint}[2][]{\url{#2}}

\bibitem{Jaeger:96}
Jaeger H~M, Nagel S~R and Behringer R~P 1996 {\em Rev. Mod. Phys.\/} {\bf
  68} 1259--1273

\bibitem{Aranson:06}
Aranson I~S and Tsimring L~S 2006 {\em Rev. Mod. Phys.\/} {\bf 78} 641--692

\bibitem{Goldhirsch:93}
Goldhirsch I and Zanetti G 1993 {\em Phys. Rev. Lett.\/} {\bf 70}
  1619--1622

\bibitem{Li:03}
Li J, Aranson I~S, Kwok W~K and Tsimring L~S 2003 {\em Phys. Rev. Lett.\/} {\bf
  90} 134301

\bibitem{Corwin:05}
Corwin E~I, Jaeger H~M and Nagel S~R 2005 {\em Nature\/} {\bf 435} 1075--1078

\bibitem{Haff:83}
Haff P 1983 {\em J. Fluid Mech.\/} {\bf 134} 401--430

\bibitem{Brey:96}
Brey J~J, Ruiz-Montero M~J and Cubero D 1996 {\em Phys. Rev. E\/} {\bf 54}
  3664--3671

\bibitem{Esipov:97}
Esipov S~E and P{\"o}schel T 1997 {\em J. Stat. Phys.\/} {\bf 86} 1385--1395

\bibitem{Ben-naim:99}
Ben-Naim E, Chen S~Y, Doolen G~D and Redner S 1999 {\em Phys. Rev. Lett.\/}
  {\bf 83} 4069--4072

\bibitem{Nie:02}
Nie X, Ben-Naim E and Chen S 2002 {\em Phys. Rev. Lett.\/} {\bf 89} 204301

\bibitem{Supravat:12}
Dey S, Das D and Rajesh R 2011 {\em Europhys. Lett.\/} {\bf 93} 44001

\bibitem{Pathak:14a}
Pathak S~N, Das D and Rajesh R 2014 {\em Europhys. Lett.\/} {\bf 107} 44001

\bibitem{Pathak:14}
Pathak S~N, Jabeen Z, Das D and Rajesh R 2014 {\em Phys. Rev. Lett.\/} {\bf
  112} 038001

\bibitem{shinde2007violation}
Shinde M, Das D and Rajesh R 2007 {\em Phys. Rev. Lett.\/} {\bf 99} 234505

\bibitem{Losert:99}
Losert W, Cooper D~G~W, Delour J, Kudrolli A and Gollub J~P 1999 {\em Chaos\/}
  {\bf 9} 682--690

\bibitem{Rouyer:00}
Rouyer F and Menon N 2000 {\em Phys. Rev. Lett.\/} {\bf 85} 3676--3679

\bibitem{Aranson:02}
Aranson I~S and Olafsen J~S 2002 {\em Phys. Rev. E\/} {\bf 66} 061302

\bibitem{reis2007forcing}
Reis P~M, Ingale R~A and Shattuck M~D 2007 {\em Phys. Rev. E\/} {\bf 75} 051311

\bibitem{wang2009particle}
Wang H~Q, Feitosa K and Menon N 2009 {\em Phys. Rev. E\/} {\bf 80} 060304

\bibitem{tatsumi2009experimental}
Tatsumi S, Murayama Y, Hayakawa H and Sano M 2009 {\em J. Fluid Mech.\/} {\bf
  641} 521--539

\bibitem{Scholz:17}
Scholz C and P{\"o}schel T 2017 {\em Phys. Rev. Lett.\/} {\bf 118} 198003

\bibitem{vilquin2018shock}
Vilquin A, Kellay H and Boudet J~F 2018 {\em J. Fluid Mech.\/} {\bf 842}
  163--187

\bibitem{Moon:01}
Moon S~J, Shattuck M~D and Swift J~B 2001 {\em Phys. Rev. E\/} {\bf 64}
  031303

\bibitem{gayen2008orientational}
Gayen B and Alam M 2008 {\em Phys. Rev. Lett.\/} {\bf 100} 068002

\bibitem{gayen2011effect}
Gayen B and Alam M 2011 {\em Phys. Rev. E\/} {\bf 84} 021304

\bibitem{Cafiero:2002}
Cafiero R, Luding S and Herrmann H~J 2002 {\em Europhys. Lett.\/} {\bf 60} 854

\bibitem{Noije:98}
van Noije T and Ernst M 1998 {\em Granular Matter\/} {\bf 1} 57--64

\bibitem{Olafsen:99}
Olafsen J~S and Urbach J~S 1999 {\em Phys. Rev. E\/} {\bf 60} R2468--R2471

\bibitem{Kudrolli:00}
Kudrolli A and Henry J 2000 {\em Phys. Rev. E\/} {\bf 62} R1489--R1492

\bibitem{Blair:01}
Blair D~L and Kudrolli A 2001 {\em Phys. Rev. E\/} {\bf 64} 050301

\bibitem{Falcon:2013}
Falcon E, Bacri J~C and Laroche C 2013 {\em Europhys. Lett.\/} {\bf 103} 64004

\bibitem{grasselli2015translational}
Grasselli Y, Bossis G and Morini R 2015 {\em Eur. Phys. J. E\/} {\bf 38} 8

\bibitem{windows2013boltzmann}
Windows-Yule C and Parker D 2013 {\em Phys. Rev. E\/} {\bf 87} 022211

\bibitem{wildman2009granular}
Wildman R~D, Beecham J and Freeman T 2009 {\em Eur. Phys. J. Special Topics\/}
  {\bf 179} 5--17

\bibitem{Schmick:08}
Schmick M and Markus M 2008 {\em Phys. Rev. E\/} {\bf 78} 010302

\bibitem{hou2008velocity}
Hou M, Liu R, Zhai G, Sun Z, Lu K, Garrabos Y and Evesque P 2008 {\em
  Microgravity Sci. Technol.\/} {\bf 20} 73

\bibitem{baxter2007temperature}
Baxter G and Olafsen J 2007 {\em Granular Matter\/} {\bf 9} 135--139

\bibitem{puglisi1998clustering}
Puglisi A, Loreto V, Marconi U~M~B, Petri A and Vulpiani A 1998 {\em Phys. Rev.
  Lett.\/} {\bf 81} 3848

\bibitem{puglisi1999kinetic}
Puglisi A, Loreto V, Marconi U~M~B and Vulpiani A 1999 {\em Phys. Rev. E\/}
  {\bf 59} 5582

\bibitem{Vanzon:04}
van Zon J~S and MacKintosh F~C 2004 {\em Phys. Rev. Lett.\/} {\bf 93} 038001

\bibitem{Vanzon:05}
van Zon J~S and MacKintosh F~C 2005 {\em Phys. Rev. E\/} {\bf 72} 051301

\bibitem{rui2011velocity}
Rui L, Duan-Ming Z and Zhi-Hao L 2011 {\em Chin. Phys. Lett.\/} {\bf 28} 090506

\bibitem{Prosendas:18}
Das P, Puri S and Schwartz M 2018 {\em Granular Matter\/} {\bf 20} 15

\bibitem{kang2010granular}
Kang W, Machta J and Ben-Naim E 2010 {\em Europhys. Lett.\/} {\bf 91} 34002

\bibitem{Prasad:18}
Prasad V~V and Rajesh R 2019 {\em J. Stat. Phys.\/} {\bf 176} 1409-1433

\bibitem{Prasad:19}
Prasad V~V, Das D, Sabhapandit S and Rajesh R 2019 {\em J. Stat. Mech.\/} 063201

\bibitem{Feitosa2002}
Feitosa K and Menon N 2002 {\em Phys. Rev. Lett.\/} {\bf 88} 198301

\bibitem{barrat2002lack}
Barrat A and Trizac E 2002 {\em Granular Matter\/} {\bf 4} 57--63

\bibitem{Marconi:02}
Marconi U~M~B and Puglisi A 2002 {\em Phys. Rev. E\/} {\bf 66} 011301

\bibitem{wang:2003}
Wang H~q, Jin G~j and Ma Y~q 2003 {\em Phys. Rev. E\/} {\bf 68} 031301

\bibitem{Wang2008}
Wang H~Q and Menon N 2008 {\em Phys. Rev. Lett.\/} {\bf 100} 158001

\bibitem{JJBrey-non-eq:2009}
Brey J~J and Ruiz-Montero M~J 2009 {\em Phys. Rev. E\/} {\bf 80} 041306

\bibitem{Uecker:2009}
Uecker H, Kranz W~T, Aspelmeier T and Zippelius A 2009 {\em Phys. Rev. E\/}
  {\bf 80} 041303

\bibitem{Baxter:03}
Baxter G and Olafsen J 2003 {\em Nature\/} {\bf 425} 680--680

\bibitem{Baxter-PRL:2007}
Baxter G~W and Olafsen J~S 2007 {\em Phys. Rev. Lett.\/} {\bf 99} 028001

\bibitem{Comb-PRE:2008}
Combs K, Olafsen J~S, Burdeau A and Viot P 2008 {\em Phys. Rev. E\/} {\bf
  78} 042301

\bibitem{Burdeau-PRE:2009}
Burdeau A and Viot P 2009 {\em Phys. Rev. E\/} {\bf 79} 061306


\bibitem{Pagnani-PRE:2002}
Pagnani R, Bettolo~Marconi U~M and Puglisi A 2002 {\em Phys. Rev. E\/} {\bf
  66} 051304

\bibitem{Barrat-PRE:2002}
Barrat A and Trizac E 2002 {\em Phys. Rev. E\/} {\bf 66} 051303


\bibitem{Bobylev:00}
Bobylev A~V, Carrillo J~A and Gamba I~M 2000 {\em J. Stat. Phys.\/} {\bf 98}
  743--773

\bibitem{Ben-naim:00}
Ben-Naim E and Krapivsky P~L 2000 {\em Phys. Rev. E\/} {\bf 61} R5--R8

\bibitem{Baldassarri:02}
Baldassarri A, Marconi U~M~B and Puglisi A 2002 {\em Europhys. Lett.\/} {\bf
  58} 14

\bibitem{Ernst:02_a}
Ernst M~H and Brito R 2002 {\em Europhys. Lett.\/} {\bf 58} 182

\bibitem{Ernst:02}
Ernst M~H and Brito R 2002 {\em Phys. Rev. E\/} {\bf 65} 040301

\bibitem{Krapivsky:02}
Krapivsky P~L and Ben-Naim E 2002 {\em J. Phys. A\/} {\bf 35} L147

\bibitem{Ben-naim:02}
Ben-Naim E and Krapivsky P~L 2002 {\em Phys. Rev. E\/} {\bf 66} 011309

\bibitem{Antal:02}
Antal T, Droz M and Lipowski A 2002 {\em Phys. Rev. E\/} {\bf 66} 062301

\bibitem{Santos:03}
Santos A and Ernst M~H 2003 {\em Phys. Rev. E\/} {\bf 68} 011305

\bibitem{Prasad:13}
Prasad V~V, Sabhapandit S and Dhar A 2013 {\em Europhys. Lett.\/} {\bf 104}
  54003

\bibitem{Prasad:14}
Prasad V~V, Sabhapandit S and Dhar A 2014 {\em Phys. Rev. E\/} {\bf 90}
  062130

\bibitem{Prasad:17}
Prasad V~V, Das D, Sabhapandit S and Rajesh R 2017 {\em Phys. Rev. E\/} {\bf
  95} 032909

\end{thebibliography}
\providecommand{\newblock}{}

\end{document}